\documentclass[11pt,preprint]{aastex}

\slugcomment{Accepted to the Astrophysical Journal Supplement Series}

\newcommand{\uv}{\mbox{$u$-$v$}}
\newcommand{\GPB}{\mbox{\em GP-B}}
\newcommand{\kms}{\mbox{km s$^{-1}$}}
\newcommand{\Jb}{\mbox{Jy bm$^{-1}$}}
\newcommand{\Msol}{\mbox{$M$\raisebox{-.6ex}{$\odot$}}}
\newcommand{\Rsol}{\mbox{$R$\raisebox{-.6ex}{$\odot$}}}

\newcommand{\thelg}{\mbox{$\theta_b$}}
\newcommand{\psvb}{\mbox{\bf b}}
\newcommand{\invd}{\mbox{d$^{-1}$}}
\newcommand{\HR}{\mbox{IM Peg}}

\shortauthors{Bietenholz et al.}
\shorttitle{{\em GP-B} VII: Radio Structure of IM Pegasi}

\defcitealias{GPB-I}{Paper I}
\defcitealias{GPB-II}{Paper II}
\defcitealias{GPB-III}{Paper III}
\defcitealias{GPB-IV}{Paper IV}
\defcitealias{GPB-V}{Paper V}
\defcitealias{GPB-VI}{Paper VI}
\defcitealias{GPB-VII}{Paper VII}

\begin{document}
      
\title{VLBI for {\em Gravity Probe B}. VII. The Evolution of the Radio Structure of IM Pegasi}

\author{M. F. Bietenholz\altaffilmark{1,2}, N. Bartel\altaffilmark{1},
D. E. Lebach\altaffilmark{3}, R. R. Ransom\altaffilmark{1,4},
M. I. Ratner\altaffilmark{3}, and I. I. Shapiro\altaffilmark{3}}

\altaffiltext{1}{Department of Physics and Astronomy, York University,
4700 Keele Street, Toronto, ON M3J 1P3, Canada}

\altaffiltext{2}{Now also at Hartebeesthoek Radio Astronomy Observatory,
PO Box 443, Krugersdorp 1740, South Africa}

\altaffiltext{3}{Harvard-Smithsonian Center for Astrophysics, 60
Garden Street, Cambridge, MA 02138, USA}

\altaffiltext{4}{Now at Okanagan College, 583 Duncan Avenue West,
  Penticton, B.C., V2A 2K8, Canada and also at the National Research
  Council of Canada, Herzberg Institute of Astrophysics, Dominion
  Radio Astrophysical Observatory, P.O. Box 248, Penticton, B.C., V2A
  6K3, Canada}



\begin{abstract}
We present measurements of the total radio flux density as well as
very-long-baseline interferometry (VLBI) images of the star,
IM~Pegasi, which was used as the guide star for the NASA/Stanford
relativity mission Gravity Probe B\@.  We obtained flux densities and
images from 35 sessions of observations at 8.4~GHz ($\lambda =
3.6$~cm) between 1997 January and 2005 July.  The observations were
accurately phase-referenced to several extragalactic reference
sources, and we present the images in a star-centered frame, aligned
by the position of the star as derived from our fits to its orbital
motion, parallax, and proper motion.  Both the flux density and the
morphology of \HR\ are variable.  For most sessions, the emission
region has a single-peaked structure, but 25\% of the time, we
observed a two-peaked (and on one occasion perhaps a three-peaked)
structure.  On average, the emission region is elongated by $1.4 \pm
0.4$~mas (FWHM), with the average direction of elongation being close
to that of the sky projection of the orbit normal.  The average length
of the emission region is approximately equal to the diameter of the
primary star.  No significant correlation with the orbital phase is
found for either the flux density or the direction of elongation, and
no preference for any particular longitude on the star is shown by the
emission region.
\end{abstract}

\section{Introduction}
\label{sintro}

\objectname{IM Pegasi} (\objectname{HR 8703}, \objectname{HD 216489},
\objectname{FK5 3829}; \HR\ hereafter) is a close binary RS~Canum
Venaticorum (RS~CVn) star \citep[]{Hall1976}.  
This star was chosen as the reference star for Gravity Probe~B (\GPB),
the spaceborne relativity experiment developed by NASA and Stanford
University; for an introduction to \GPB, see \citet{GPB-I}, hereafter
Paper~I\@.  The \GPB\ experiment required a very accurate measurement
of the proper motion of the guide star, which was determined by using
phase-referenced very-long-baseline interferometry (VLBI) measurements
relative to the compact extragalactic objects \objectname[]{3C 454.3},
\objectname[]{QSO B2250+194}, and B2252+172.
This paper is the seventh and last in a series describing the \GPB\
experiment and the astrometric observations carried out in its
support.  The first, mentioned above (Paper~I), contains a general
introduction.  The radio structure and astrometric stability of the
reference sources are described in Papers~II \citep{GPB-II} and III
\citep{GPB-III}, respectively.  The astrometric process and results
are described in Paper~IV \citep{GPB-IV}, Paper~V \citep{GPB-V}, and
Paper~VI \citep{GPB-VI}.  Early results on radio imaging and
astrometry of \HR\ were presented in \citet{Ransom-VLBA10th}, and a
brief overview was given in \citet{NB-GPB2008}.

In this paper, we discuss the unprecedented sequence of VLBI images of
\HR\ obtained as a result of our VLBI observations.  These images,
which have resolutions (east-west) typically about equal to the
stellar angular radius, were obtained in addition to the astrometry
required for the \GPB\ project.  This set of images is the most
extensive available for any radio star 
\footnote{We note that \citet{Peterson+2011} have made VLBI 
observations of the Algol and UX Arietis systems with slightly fewer
epochs per source, but with coverage over longer spans of time.}.
RS~CVn stars are known to have strong and relatively compact radio
emission, and to be variable in radio flux density on time scales of
an hour or even less \citep[e.g.,][]{Lebach+1999, Jones+1996,
Hjellming1988, Dulk1985, Mutel+1985}. For other VLBI images of RS~CVn
stars, we refer the interested reader to \citep{Ransom+2003,
Ransom+2002, Lestrade+1995, Massi+1988, Mutel+1985}.

The guide star \HR\ is at a distance of $96.4 \pm 0.7$~pc (from VLBI
parallax measurement; Hipparcos optical parallax measurements by
ESA 1997 gave a consistent measure of this distance,
albeit with rather larger uncertainties; see Paper
V)\nocite{PerrymanE1997}\nocite{GPB-V}.  
The star has an orbital period of $\sim$25~d.  The primary is a K2~III
star with a mass of $1.8 \pm 0.2$~\Msol\ and an effective temperature
ranging from $\sim$3500 to 5100~K \citep{Berdyugina+2000}. It rotates
rapidly, with $v \sin{i} \sim 27$~\kms\ \citep{Berdyugina+2000}, and
its radius is $13.3 \pm 0.6$~\Rsol\ \citep{BerdyuginaIT1999a}, so its
angular diameter would be $1.28 \pm 0.06$~mas on the sky.  The
secondary, first spectroscopically detected by \citet{Marsden+2005},
has a mass of $\sim$1~\Msol, and an effective temperature of
$\sim$5650~K\@.  The orbit has been accurately determined using
optical spectroscopy, and we use here the parameters given in
\nocite{Marsden+2005} Marsden et al.\ \citep[2005; see
  also][]{BerdyuginaIT1999a}, namely a period of $24.64877 \pm
0.00003$~d, with a (heliocentric) time for superior conjunction of the
primary, with the primary at maximum distance from us, being the
Julian date $2,450,342.905 \pm 0.004$, and an orbital
eccentricity\footnote{\citet{BerdyuginaM2006} suggest that the orbital
  eccentricity is 0.017; such a small deviation from 0 would not
  affect any of our conclusions.} of 0\@.  The inclination angle of
the orbit to the plane of the sky as determined from our fit of the
orbit to the positions determined by VLBI is $73\arcdeg \pm 8\arcdeg$
and the p.a.\ of the ascending node is $41 \pm 9\arcdeg$
\citepalias{GPB-V}.  This value is consistent with the earlier lower
limit of 55\arcdeg\ set by \citet{Lebach+1999}, and is in good
agreement with the range of 65\arcdeg\ and 80\arcdeg\ determined from
optical observations \citep{BerdyuginaIT1999b}.

In the optical, brightness variations of $\sim$0.3~mag have been
observed \citep[Paper I;][]{RibarikOS2003, Strassmeier+1997}.  Doppler
imaging and photometry have shown that there are dark spots, which
cover $\sim$15\% of the star's surface, with temperatures more than
1500~K lower than the average.  These spots persist over several
orbital periods, although they seem to drift slowly on the surface of
the star with respect to its orbital phase \citep{Zellem+2010,
BerdyuginaM2006, RibarikOS2003, Berdyugina+2000}

\section{Observations and Data Reduction}
\label{sobs}

The VLBI observations of \HR\ were carried out at 35 observing
sessions between 1997 and 2005, using a global array of radio
telescopes, at a frequency of 8.4~GHz.  A fuller description of the
VLBI observations as well as the basic data reduction is given in
\citetalias{GPB-II}.

The National Radio Astronomy Observatory's phased Very Large Array
(VLA) took part in 32 of the 35 sets of VLBI observations.  By using
the interferometric data from the VLA, we obtained accurate total
flux-density measurements during the VLBI observations.  We reduced
the VLA data following standard procedures using NRAO's AIPS software
package, with the amplitudes calibrated by using observations of the
standard flux-density calibrators (3C~286 and 3C~48) and the scale of
\citet{Baars+1977}.  The flux densities were determined from images
made from data which we self-calibrated in phase but not in amplitude.
Since, for the majority of our observing sessions, the flux density
varied significantly during the session, we list in
Table~\ref{tepochs} the maximum and minimum flux density during each
observing session as estimated from the lightcurves, along with the
corresponding calendar date and Modified Julian date (MJD) of the
midpoint of the observing session.  We also list the fractional
circular polarization, $m_c$.  The circular polarization was
calibrated by assuming that the calibrator sources have $m_c = 0$, and
by ignoring the leakage terms in the polarization response of the VLA
antennas.  The first of these assumptions is generally true to better
than 1\% \citep{RaynerNS2000}, and the effect of the leakage terms at
the VLA also has an effect of $<1$\% on the derived values of $m_c$.
For the session of 2004 March 6, we performed a full polarization
calibration, and determined the leakage terms. We found that \HR\ had
no detectable linear polarization.

\begin{deluxetable}{l@{\hspace{0.2in}}c c l l r}
\tabletypesize{\tiny}
\tablecaption{Observing Sessions and Radio Flux Densities at 8.4 GHz}

\tablehead{
\colhead{Date} & \colhead{MJD} & \colhead{Orbital} &
          \multicolumn{2}{c}{Total flux density\tablenotemark{b}} & \colhead{Fractional circular} \\
               &               & \colhead{phase\tablenotemark{a}}   
                                                   & \colhead{Minimum} & \colhead{Maximum} &
\colhead{Polarization, $m_c$\tablenotemark{c}} \\
               &               &                   & \hspace{5pt}(mJy) & \hspace{5pt}(mJy)   & (\%)\hspace{10pt} 
}
\startdata
 1997 01 16 & 50464.90 & 0.97 &   18   &   46   &$ -0.5\pm2.0$ \\ 
 1997 01 18 & 50466.89 & 0.05 &\phn6.8 &   21   &$ -1.5\pm2.0$ \\ 
 1997 11 30 & 50782.03 & 0.84 &\phn8.5 &   13   &$ -1.4\pm2.0$ \\ 
 1997 12 21 & 50803.96 & 0.73 &   48   &   76   &$ -2.7\pm2.0$ \\ 
 1997 12 27 & 50809.96 & 0.97 &\phn8.4 &   18   &$  2.9\pm2.0$ \\ 
 1998 03 01 & 50873.78 & 0.56 &\phn1.1 &   24   &$  3.6\pm2.0$ \\ 
 1998 07 12 & 51006.41 & 0.94 &\phn1.8 &\phn2.4 &$ -3.7\pm2.0$ \\ 
 1998 08 08 & 51033.35 & 0.03 &\phn2.9 &   16   &$  1.9\pm2.0$ \\ 
 1998 09 17 & 51073.24 & 0.65 &   11   &   28   &$  3.6\pm2.0$ \\ 
 1999 03 13 & 51250.74 & 0.85 &\phn0.7 &\phn3.3 &$  0.9\pm2.0$ \\ 
 1999 05 15 & 51313.57 & 0.40 &\phn2.0 &\phn7.3 &$  4.7\pm2.0$ \\ 
 1999 09 19 & 51440.23 & 0.54 & 12     &   25   &$ -0.9\pm2.0$ \\ 
 1999 12 09 & 51521.99 & 0.86 &\phn2.1 &\phn3.8 &$  1.2\pm2.0$ \\ 
 2000 05 15 & 51679.56 & 0.25 &\phn0.2 &\phn0.9 &$ 12.3\pm3.7$ \\ 
 2000 08 07 & 51763.34 & 0.65 &\phn8.0 &   58   &$  0.3\pm2.0$ \\ 
 2000 11 06 & 51854.09 & 0.33 &\phn0.6 &\phn5.7 &$  2.0\pm2.0$ \\ 
 2000 11 07 & 51855.01 & 0.37 &\phn8.4 &  10.6  &$ -0.5\pm2.0$ \\ 
 2001 03 31 & 51999.73 & 0.24 &\phn0.3 &\phn0.3 &$ 10.4\pm9.7$ \\ 
 2001 06 29 & 52089.48 & 0.88 &\phn0.4 &\phn1.0 &$-20.1\pm4.0$ \\ 
 2001 10 20 & 52202.05 & 0.45 &\phn4.2 &\phn8.1 &$ -4.3\pm2.0$ \\ 
 2001 12 21 & 52264.99 & 1.00 &\phn1.2 &\phn1.2 &$  5.5\pm2.1$ \\ 
 2002 04 14 & 52378.65 & 0.61 &\phn0.3 &\phn0.6 &$ 16.3\pm3.6$ \\ 
 2002 07 14 & 52469.40 & 0.29 &\phn0.47\tablenotemark{d}  
                                     &\phn0.47   \\ 
 2002 11 21 & 52599.06 & 0.55 &\phn0.32\tablenotemark{d}   
                                     &\phn0.32   \\ 
 2003 01 26 & 52665.88 & 0.26 &\phn0.20&\phn0.28 &$ 1.0\pm5.8$ \\ 
 2003 05 18 & 52777.55 & 0.79 &\phn1.0 &\phn1.1 &$ 15.0\pm5.0$ \\ 
 2003 09 09 & 52891.24 & 0.41 &\phn0.5 &\phn0.7 &$ 13.3\pm3.2$ \\ 
 2003 12 06 & 52979.00 & 0.97 &\phn0.9\tablenotemark{d}  
                                     &\phn0.9    \\ 
 2004 03 06 & 53070.76 & 0.69 &\phn8.0  & 19    &$ -0.5\pm2.0$ \\ 
 2004 05 18 & 53143.58 & 0.64 & 10      & 12    &$ -1.4\pm2.0$ \\ 
 2004 06 26 & 53182.49 & 0.22 &\phn5.0  & 12    &$  0.9\pm2.0$ \\ 
 2004 12 12 & 53351.00 & 0.06 &\phn0.7  &\phn1.1  &$ -6.3\pm2.9$ \\ 
 2005 01 15 & 53385.92 & 0.48 &\phn0.25 &\phn0.35 &$ 44.3\pm5.9$ \\ 
 2005 05 28 & 53518.45 & 0.86 &\phn0.25 &\phn0.85 &$ 27.9\pm5.5$ \\ 
 2005 07 16 & 53567.41 & 0.84 &\phn0.24 &\phn0.36 &$ 44.6\pm7.7$ \\  
\enddata
\tablenotetext{a}{The orbital phase for the midpoint of our observing session, in 
fractions of an orbit period, as determined from the orbit of \citet{Marsden+2005}}.
\tablenotetext{b}{Minimum and maximum total radio flux densities during the
observing session, estimated from the lightcurves from concurrent VLA
measurements, after boxcar smoothing with a window about 20~min wide,
except for the noted sessions where the flux densities were estimated
from the VLBI measurements.  The measurement uncertainties in
the flux densities are usually dominated by the systematic uncertainty
in the flux-density calibration which we take to be 5\% for the VLA
and 10\% for VLBI\@.  We do not list the uncertainties because the
intrinsic variability of \HR\ on short time scales is generally larger
than the measurement uncertainty.}
\tablenotetext{c}{Average of the fractional circular polarization,
$m_c$, (labeled FCP; IEEE convention) during the observing session, in
\%.  We adopt a minimum $1 \sigma$ uncertainty in the circular
polarization of 2\%.}
\tablenotetext{d}{Flux densities determined from VLBI measurements.}
\label{tepochs}
\end{deluxetable}

Turning again to the VLBI data, we used phase-referenced astrometry,
as described in \citetalias{GPB-IV} and V, to determine with
unprecedented accuracy the proper motion of \HR.  In particular, we
estimated separately the star's secular proper motion, its projected
orbit, and its parallax, all with respect to extragalactic reference
sources.  Knowledge of this proper motion and parallax allows us to
place all the VLBI images of \HR\ in a star-centered frame.  In
particular, our solution for the overall motion of \HR\ allows us to
estimate the position of the star's center for each observing session
with a $1 \sigma$ uncertainty of $\lesssim 0.30$~mas in an
extragalactic frame \citepalias{GPB-V}.

Our final images were phase referenced to a near-stationary feature in
the brightness distribution of the quasar 3C~454.3.  As the astrometry
was expected to be critical for the success of the \GPB\ project, it
was done more elaborately than usual.  We used both phase-referenced
mapping and parametric model fits to the measured fringe phases using
a Kalman filter.  The fringe and phase calibration used for the final
images in this paper is the same as was used to obtain our final
astrometric results, and involved a combination of both these methods.
The two methods and their relative advantages, as well as a new
combination of the two, are described in \citetalias{GPB-IV}\@.  We
carried out amplitude calibration in AIPS, using CLEAN models of the
calibrator sources, 3C~474.3, B2250+194, and B2252+172 (see Papers II
and III).  The resulting phase-referenced images of \HR\ should
represent a relatively unbiased estimate of the source brightness
distribution, since the calibration was derived without any reference
to the \HR\ visibilities.  In particular, no spurious symmetrization
should have occurred, as often results from self-calibration that is
based on an initial point-source model \citep[see][]{MassiA1999}, nor
should any reduction of the circular polarization have taken place, as
can occur when one performs amplitude self-calibration.  We note that
as with the VLA data above, we have assumed that our calibrator
sources had $m_c$ = 0, which assumption is unlikely to introduce errors
of $>1$\% in our circular polarization images of \HR.  Furthermore,
any motion of the radio emission during the course of an observing
session has not been significantly suppressed by self-calibration in
phase.

\section{Results}
\label{sresults}

\subsection{Total Flux Densities}
\label{stotflux}

\begin{figure}
\centering
\includegraphics[width=0.48\textwidth,trim=0 0.1in 0 0.4in,clip]{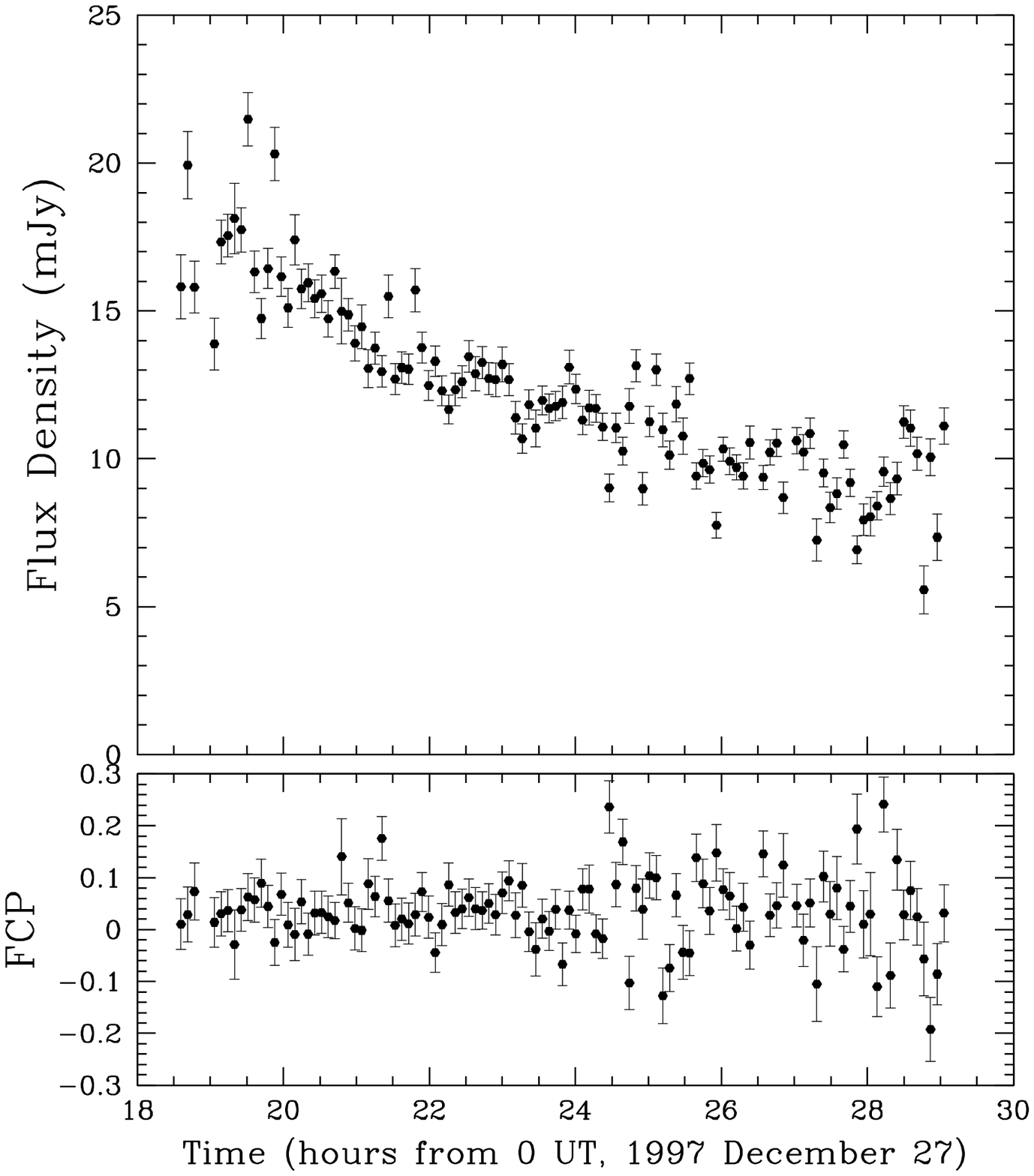}
\includegraphics[width=0.48\textwidth,trim=0 0.1in 0 0.4in,clip]{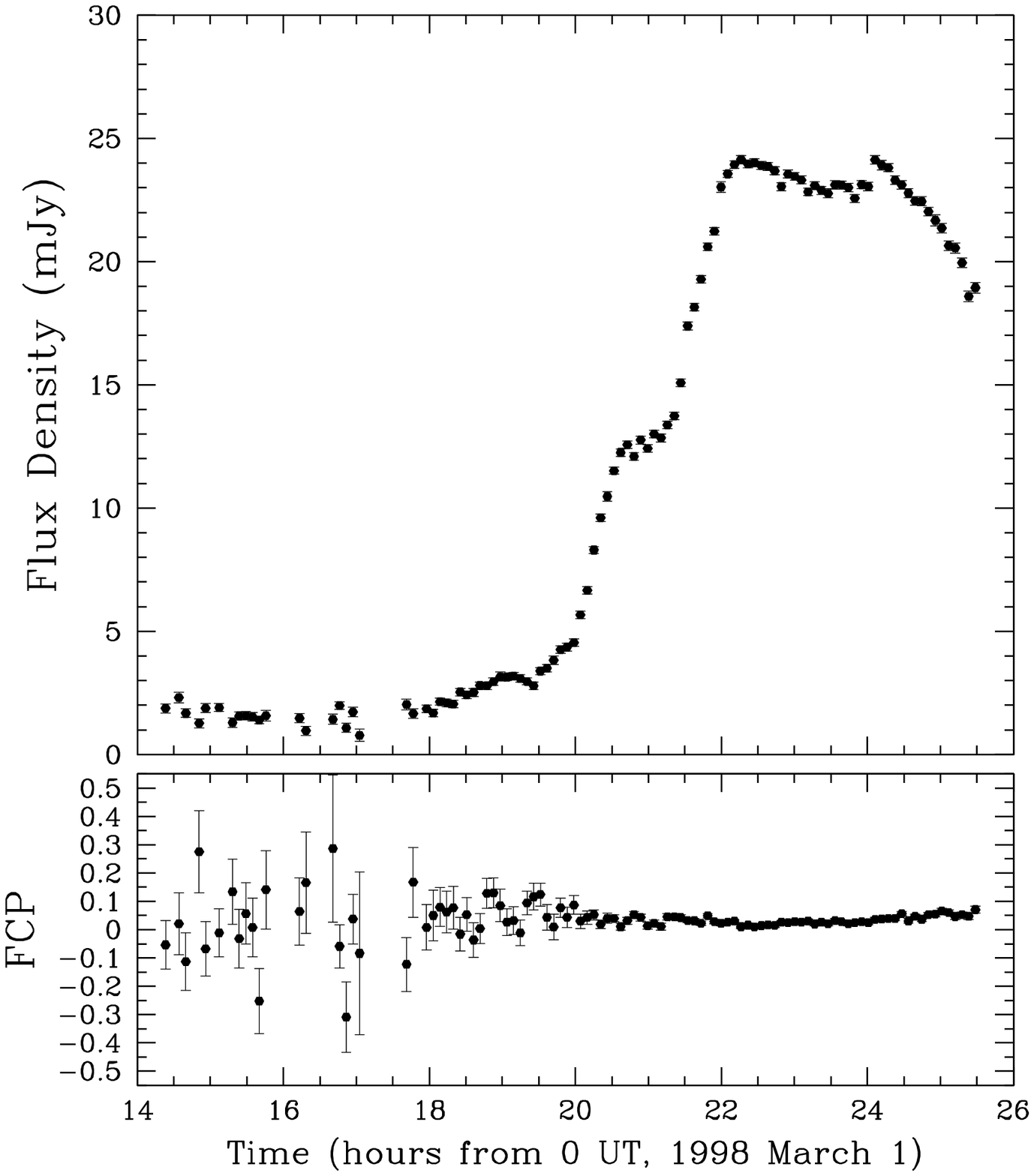}
\includegraphics[width=0.48\textwidth,trim=0 0.1in 0 0.4in,clip]{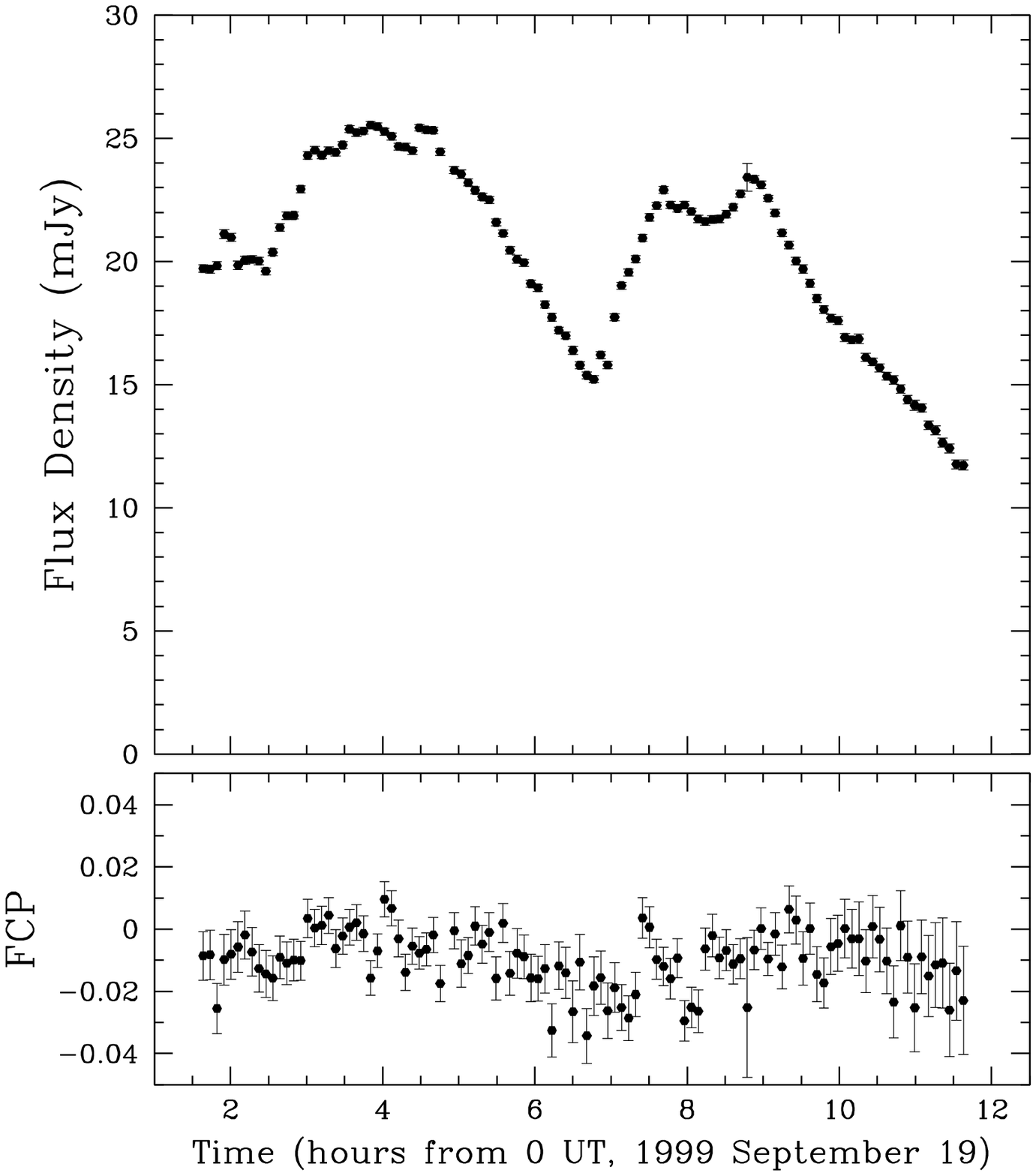}
\includegraphics[width=0.48\textwidth,trim=0 0.1in 0 0.4in,clip]{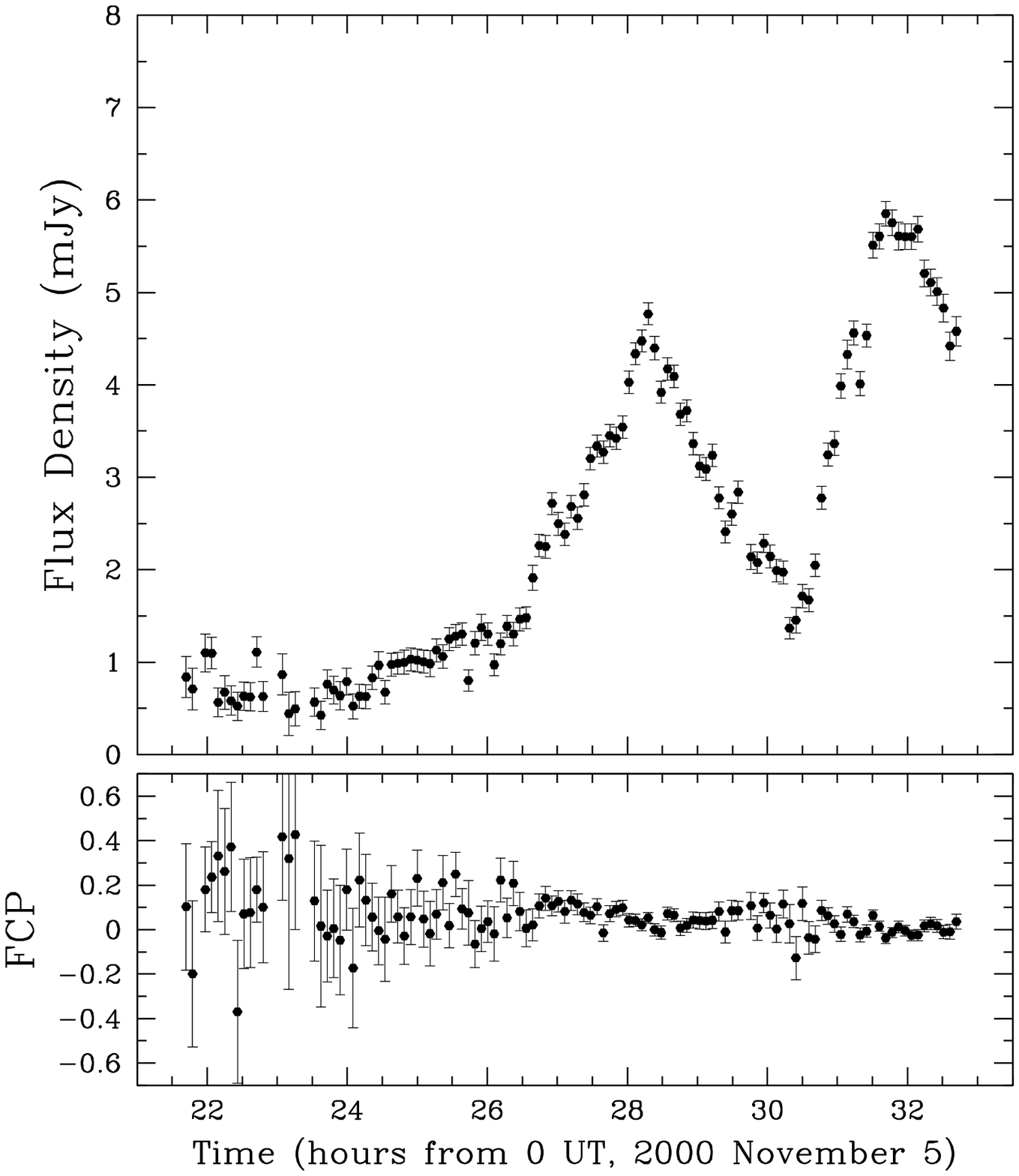}
\caption{The total 8.4-GHz flux density (top curves) and fractional
circular polarization, $m_c$, (IEEE convention; lower curves) of \HR\
as functions of time, as determined from VLA observations, shown for
four example observing sessions, namely 1997 Dec 27, 1998 Mar 1, 1999
Sep 19 and 2000 Nov 6.
Note that the dates given in horizontal axis labels in this figure are
start dates, so that the lower right panel refers to the observing
session with the midpoint date of 2000 Nov 6 in Table~\ref{tepochs}.
The plotted $1 \sigma$ uncertainties are statistical only and do not
include an estimated 5\% systematic uncertainty in the VLA
flux-density calibration. (Within each session, any calibration error
in the flux density is not expected to vary rapidly as a function of
time.)}
\label{flightcurves}
\end{figure}

We determined the total radio flux density of \HR\ during each of our
observing sessions, in most cases from VLA observations.  \HR's flux
density varied detectably during almost all observing sessions, with
variations on time scales as short as $\sim$30~min, and over a range in
flux density exceeding 30:1 in a single session (and overall exceeding
300:1).  We tabulate the maximum and minimum flux densities for each
session in Table~\ref{tepochs}, and show four example lightcurves in
Figure~\ref{flightcurves}.

Significant circular polarization, also generally time-variable, was
observed in many cases, and we list as a percentage the average
circular polarization, defined as $(R - L)/(R+L)$,
for each observing session in Table~\ref{tepochs}.  We defer
discussion of the circular polarization to \S~\ref{spol} below.

\subsection{VLBI Images} 
\label{simages}

\begin{figure}[tp]
\centering
\includegraphics[width=0.93\textwidth]{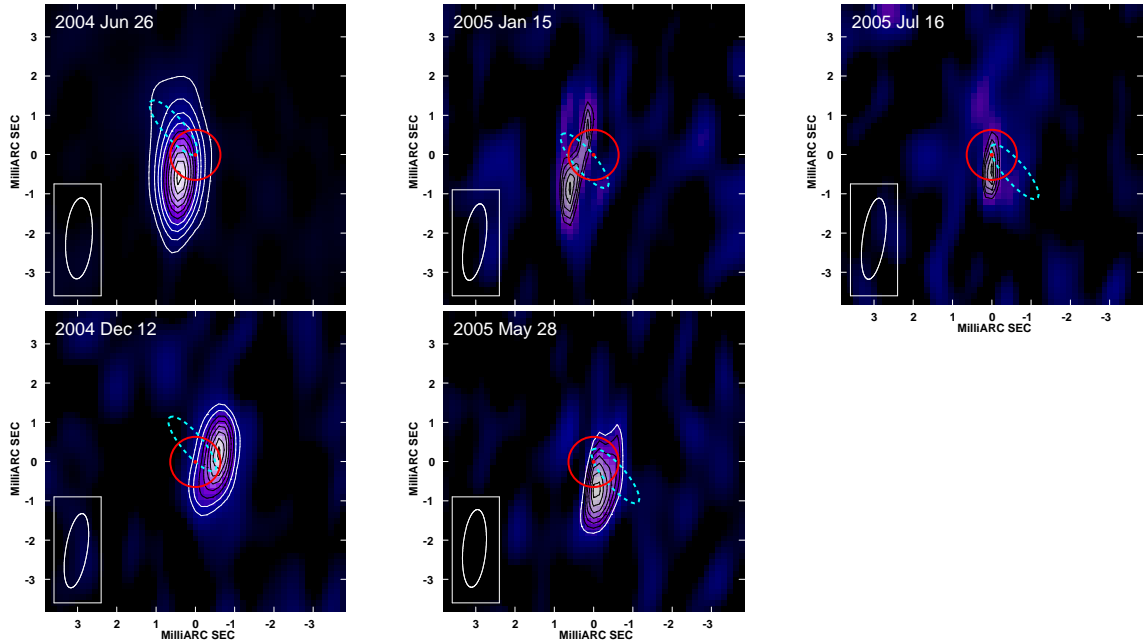}
\caption{The VLBI images of \HR\ at 8.4~GHz.  The observing date is
indicated in the top left of each panel, while the FWHM of the
convolving beam is indicated in the lower left.  North is up and east to
the left. Both the contours and the color scale show the brightness.
The contours are drawn at 10, 20, 30, \dots, 80, 90, and 98\% of the
peak brightness, starting with the first contour above $3\times$ the
rms background; and contours at 50\% and above are drawn in black.
The peak brightness and rms background values are listed in
Table~\ref{timages}.  The center of each panel (red dot) is the fit
position of the star's center, as derived from the astrometric
results of \citepalias{GPB-V}.  In other words, the coordinate origin
in our radio image should approximately represent the center of the
disk of
the primary star.  The red circle indicates the angular size of the
primary star
\citep[radius of $13.3 \pm 0.6$~\protect\Rsol;][]{BerdyuginaIT1999a}.
The cyan dotted ellipse shows the binary orbit of the primary
\citepalias{GPB-V}.  Note that we have chosen to keep the primary
star, rather than the center of the binary orbit at the center of our
plots.
For the on-line edition, we include also an mpeg animation showing the
evolution of \HR's radio emission.  This animation consists of a
simple linear interpolation in brightness between consecutive observing
sessions.  It is intended to be illustrative only, as our sampling is
not rapid enough to follow the evolution of the radio emission in
detail.}
\label{fimages}
\end{figure}

In Figure \ref{fimages}, we present the 8.4~GHz VLBI images of \HR\
for all 35 observing sessions, while Table~\ref{timages} lists some
characteristics of the images including their peak and rms background
brightness values.  As shown in Table~\ref{tepochs}, \HR's total flux
density, as monitored by VLA observations, varied significantly during
a number of our VLBI observing sessions.  In such cases, the VLBI
image made from the whole data set (such as those in
Fig.~\ref{fimages}) represents the average emission over the observing
session, albeit with some possible distortion due to the incomplete
removal of sidelobes\footnote{Images made from earth-rotation aperture
synthesis observations, such as our VLBI images, are generally based
on the assumption that the source brightness distribution does not
change during the observations.  Since both the brightness
distribution and the instantaneous point-spread function are time
variable, and since deconvolution is a non-linear process, the
deconvolved image does not strictly represent the time-averaged
brightness distribution.  However, provided that the change in the
brightness distribution is not gross and that the sidelobes are
relatively low compared to the peak of the point-spread function, a
deconvolved image should be a reasonable approximation to the
time-averaged brightness distribution.}.
The rms background brightness values, given in Table~\ref{timages},
are estimated from empty regions in each image. They represent a lower
bound on the image brightness uncertainty.  Adding to this bound are
the effects of the aforementioned distortion due to time-variability,
and deconvolution errors, which limit the image fidelity (see, e.g.,
\citealt{BriggsSS1999} and \citealt{Briggs1995}; see also discussion
in \citealt{SN93J-3}.  The images were made by first weighting
the visibilities with the inverse square root of the nominal
statistical variance in the visibility measurements\footnote{The
nominal statistical variance of the visibility measurements reflects
the uncertainties due only to random measurement noise.  If such
random noise were the only source of error in the visibility
measurements, then the most efficient weighting would be weighting by
the inverse of the nominal variance.  However, in our case, especially
for the more sensitive telescopes, the effective uncertainty in a
particular visibility measurement can be dominated by small residual
calibration errors rather than statistical noise.  These calibration
errors will depend only weakly on the telescope sensitivity.  For this
reason, we weighted by the inverse square root of the variance.  This
compresses the weights, and although \emph{ad hoc}, should lead to a
more robust image at the possible expense of a slight loss of
signal-to-noise ratio.}
and then further modifying the weighting by the robust weighting
scheme of \citet{Briggs1995} as implemented in AIPS.

\begin{deluxetable}{l@{\hspace{0.2in}}c c c c c c}
\tabletypesize{\scriptsize}
\tablecaption{Image Characteristics}

\tablehead{
\colhead{Date} & \colhead{MJD} & \colhead{Peak brightness} & \colhead{Image RMS\tablenotemark{a}} 
& Major axis\tablenotemark{b,e} &  Axis ratio \tablenotemark{c,e} & P.A.\tablenotemark{d,e}\\
(midpoint) & (midpoint)         &   (m\Jb)                  &   (m\Jb) 
& (mas)        &       & (\arcdeg) 
}
\startdata
 1997 01 16 & 50464.90 & \phn 7.97 &  0.15 & 2.78 & 0.22 & 136 \\     
 1997 01 18 & 50466.89 & \phn 4.67 &  0.07 & 1.93 & 0.40 & 149 \\     
 1997 11 30 & 50782.03 & \phn 6.85 &  0.06 & 1.00 & 0.57 & 140 \\     
 1997 12 21 & 50803.96 & 31.9 \phn &  0.15 & 0.77 & 0.79 &\phn13 \\   
 1997 12 27 & 50809.96 & \phn 4.61 &  0.06 & 2.18 & 0.27 &  146 \\    
 1998 03 01 & 50873.78 & \phn 5.52 &  0.22 & 1.49 & 0.00 &  124 \\    
 1998 07 12 & 51006.41 & \phn 0.72 &  0.04 & 1.62 & 0.50 &  134 \\    
 1998 08 08 & 51033.35 & \phn 4.03 &  0.06 & 0.89 & 0.46 &  130 \\    
 1998 09 17 & 51073.24 & 12.6 \phn &  0.13 & 0.99 & 0.46 &  127 \\    
 1999 03 13 & 51250.74 & \phn 0.92 &  0.04 & 1.64 & 0.35 &  153 \\    
 1999 05 15 & 51313.57 & \phn 2.09 &  0.04 & 0.98 & 0.48 &  105 \\    
 1999 09 19 & 51440.23 & 13.1 \phn &  0.10 & 1.38 & 0.75 &  170 \\    
 1999 12 09 & 51521.99 & \phn 1.05 &  0.04 & 0.29 & 0.52 &\phn93 \\   
 2000 05 15 & 51679.56 & \phn 0.39 &  0.04 & 1.43 & 0.15 &  149 \\    
 2000 08 07 & 51763.34 & 32.7 \phn &  0.43 & 0.53 & 0.62 &  109 \\    
 2000 11 06 & 51854.09 & \phn 1.87 &  0.05 & 1.01 & 0.00 &  125 \\    
 2000 11 07 & 51855.01 & \phn 6.82 &  0.12 & 0.89 & 0.30 &\phn97 \\   
 2001 03 31 & 51999.73 & \phn 0.23 &  0.05 & 2.53 & 0.00 &\phn10 \\   
 2001 06 29 & 52089.48 & \phn 0.44 &  0.03 & 1.09 & 0.48 &\phn16 \\   
 2001 10 20 & 52202.05 & \phn 5.17 &  0.07 & 0.95 & 0.57 &  134 \\    
 2001 12 21 & 52264.99 & \phn 0.78 &  0.05 & 1.12 & 0.93 &\phn92 \\   
 2002 04 14 & 52378.65 & \phn 0.36 &  0.03 & 0.69 & 0.00 &\phn\phn3 \\
 2002 07 14 & 52469.40 & \phn 0.24 &  0.04 & 2.74 & 0.29 &\phn\phn1 \\
 2002 11 21 & 52599.06 & \phn 0.29 &  0.05 & 0.74 & 0.00 &162 \\      
 2003 01 26 & 52665.88 & \phn 0.21 &  0.05 & 0.68 & 0.00 &\phn21 \\   
 2003 05 18 & 52777.55 & \phn 0.52 &  0.04 & 0.84 & 0.85 &  160 \\    
 2003 09 09 & 52891.24 & \phn 0.24 &  0.05 & 1.46 & 0.00 &\phn69 \\   
 2003 12 06 & 52979.00 & \phn 0.27 &  0.05 & 2.41 & 0.51 &  136 \\    
 2004 03 06 & 53070.76 & \phn 9.52 &  0.09 & 0.71 & 0.54 &  179 \\    
 2004 05 18 & 53143.58 & \phn 5.28 &  0.06 & 1.02 & 0.75 &\phn28 \\   
 2004 06 26 & 53182.49 & \phn 4.77 &  0.06 & 1.13 & 0.42 &\phn\phn4 \\
 2004 12 12 & 53351.00 & \phn 0.53 &  0.04 & 1.20 & 0.65 &  177 \\    
 2005 01 15 & 53385.92 & \phn 0.14 &  0.02 & 3.3\phn & 0.22 & 175 \\ 
 2005 05 28 & 53518.45 & \phn 0.29 &  0.03 & 1.36 & 0.28 &  152 \\ 
 2005 07 16 & 53567.41 & \phn 0.13 &  0.03 & 1.50 & 0.00 & \phn\phn20 \\ 
\enddata
\tablenotetext{a}{The rms background as estimated from empty regions
in each image; this rms represents a lower limit to the image uncertainty
(see text).}
\tablenotetext{b}{The FWHM major axis, $b$, of an elliptical Gaussian
fitted to the image.}
\tablenotetext{c}{The axis ratio of an elliptical Gaussian fitted to
the image.}
\tablenotetext{d}{The p.a.\ of the major axis, \thelg, of an
elliptical Gaussian fitted to the image, constrained to be $<180\arcdeg$}
\tablenotetext{e}{The values are those for an elliptical Gaussian
which is the deconvolution of the (elliptical Gaussian) restoring beam
from the elliptical Gaussian fitted to the CLEAN image by least
squares.}
\label{timages}
\end{deluxetable}

We display the VLBI images in Fig.~\ref{fimages}, centered on the
chromospherically-active primary based on our astrometry, which
includes fits to the $\sim$25-d orbit and the parallax, as well as the
secular proper motion of the star (see Papers IV, V, and VI)\@.  This
identification of the star's center in each image is based not only on
our astrometry, but also on the assumption that the radio emission is,
on average, centered on the star.  It is possible, although we
consider it less likely, that there is a significant systematic offset
of the radio emission from the center of the star.
The rms scatter in position over all observing sessions, based on the
residuals, is $\sim$0.4~mas in both $\alpha$ and $\delta$.  This
scatter is dominated by the variability in the position of the radio
source with respect to the star's center, as our astrometric
uncertainty is $<0.14$~mas.  We expect our fit position for the star's
center, which we use as the origin in each panel of
Figure~\ref{fimages}, to be in error by no more than 0.3~mas
($1 \sigma$), with the likely errors being smaller
\citepalias[see][]{GPB-V}.

\begin{figure*}[tp]
\centering
\includegraphics[width=0.93\textwidth]{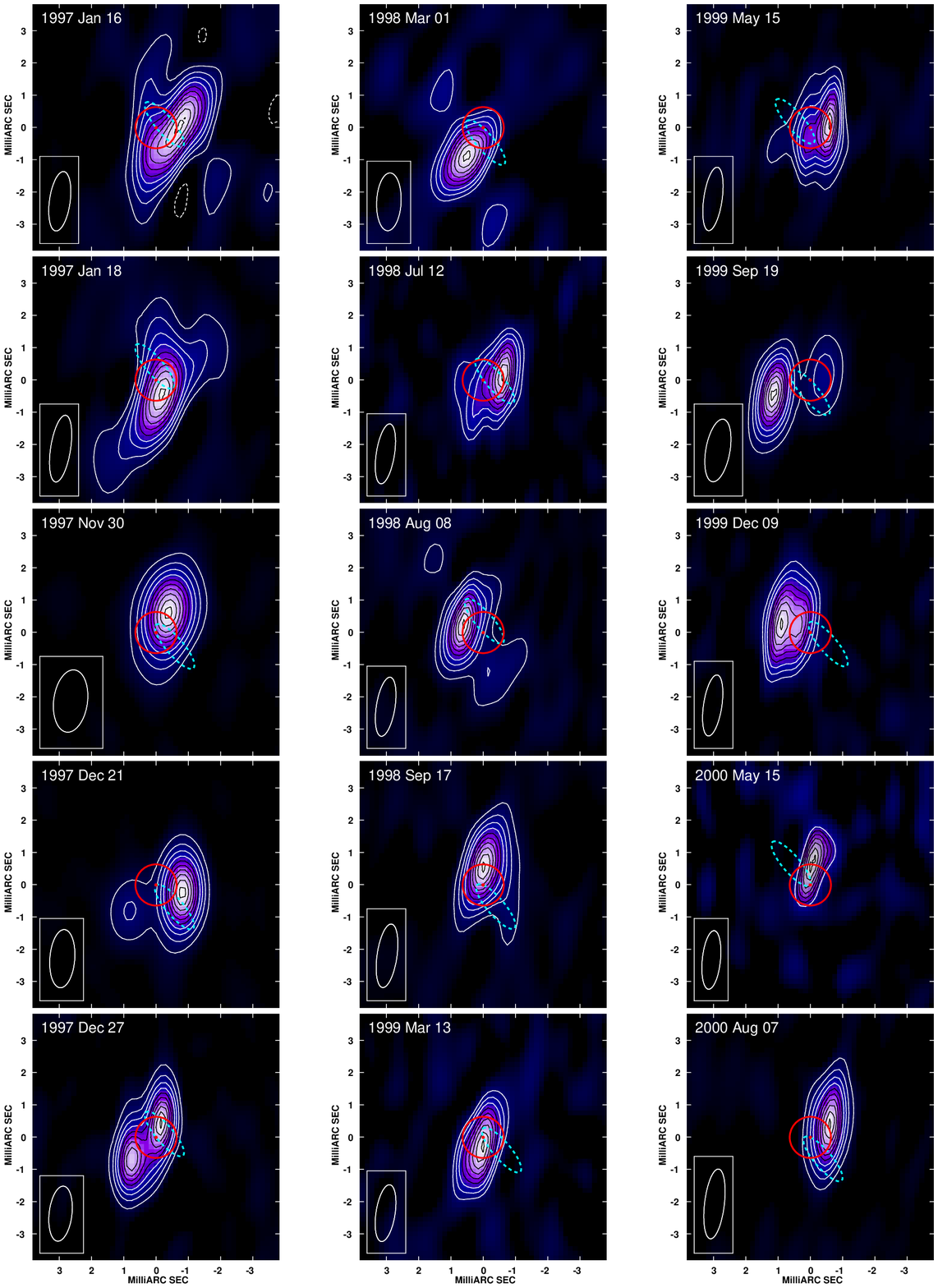}
\end{figure*}

\begin{figure*}[tp]
\centering
\includegraphics[width=0.93\textwidth]{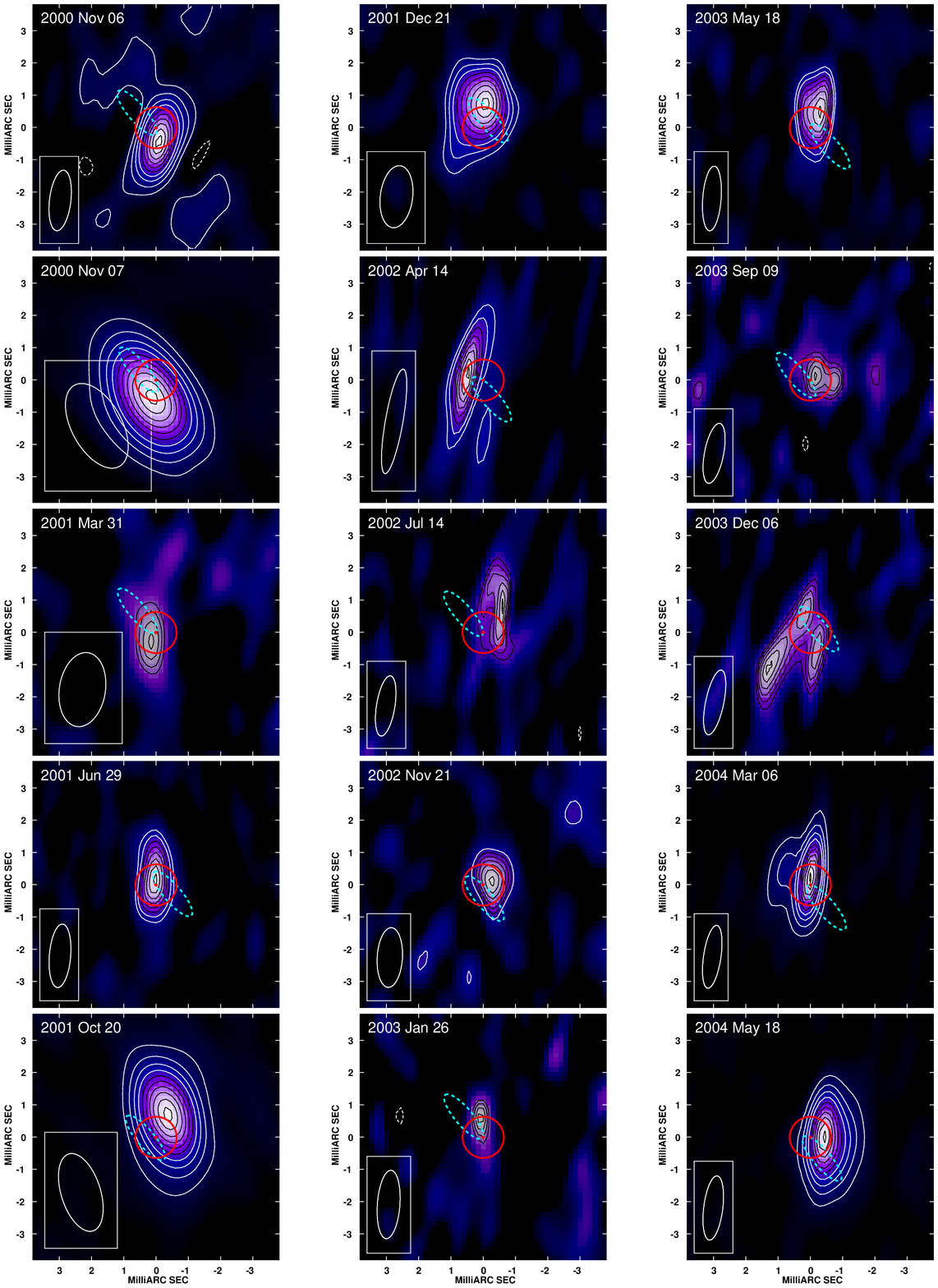}
\end{figure*}

\subsection{Rapid Time Evolution of the Images from VLBI}

\begin{figure}
\centering
\includegraphics[width=0.95\textwidth]{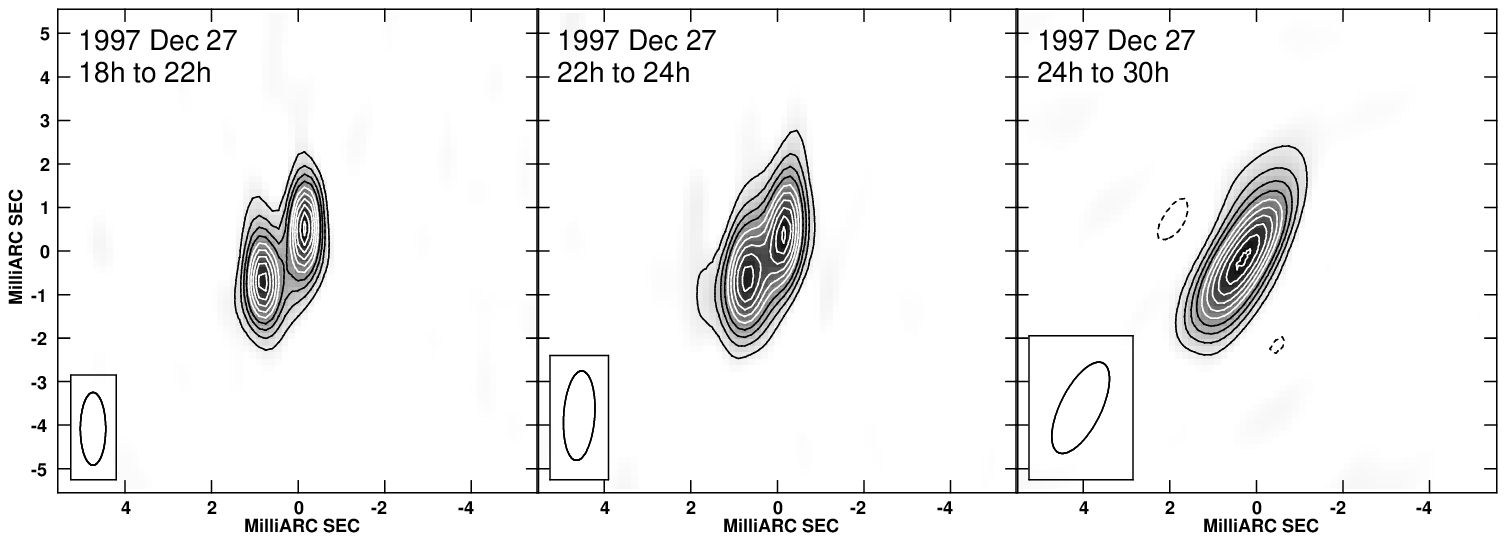}
\includegraphics[width=0.95\textwidth]{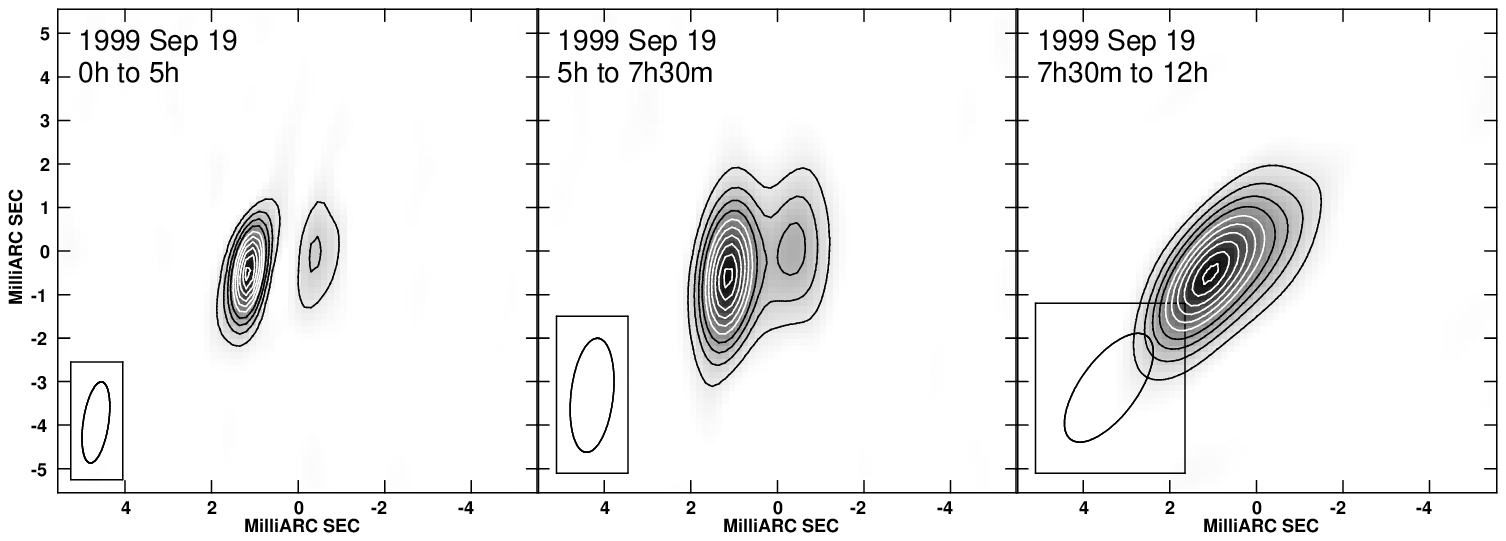}
\caption{Time-resolved images of \HR\ on 1997 December 27 (top row) and
1999 September 19 (bottom row).  
In each row, the three panels show the VLBI images in the three
consecutive UT time ranges given.  The convolving beam,
which varies with time, is indicated at lower left in each panel.
Contours and greyscale are similar to those in Fig.~\ref{fimages}.
The three time-ranges were chosen so that each has an approximately
equal number of visibility measurements. The corresponding light
curves were given in Figure~\ref{flightcurves}.
}
\label{f1997dec}
\end{figure}

The \uv~coverage of our array is dense enough to allow us to produce
images from subsets of our data for each observing session, each
covering several consecutive hours.  As two examples, we show in
Figure~\ref{f1997dec} images made by partitioning the data from the
observing sessions of 1997 December 27 and 1999 September 19,
respectively, into three time ranges.
The time ranges were chosen so as to have an approximately
equal number of visibility measurements in each interval.

Unfortunately, as the \uv~coverage changes, so does the elliptical
convolving beam.  Convolving with a common, round beam would have
allowed easier comparison of the three time ranges, but at the
expense of losing much of the image structure to the resulting
lower resolution.

Both these observing sessions show similar temporal behavior: In the
first two of the three time ranges, the source exhibits a double
structure very similar to that of the average image shown in
Fig.~\ref{fimages}, with two distinct brightness maxima.  In both
sessions, in the last of the three time ranges, the effective resolution
is lower, and it is no longer possible to distinguish the two maxima,
although the presence of a double-structure similar to that seen in
the first two time ranges is compatible with the image.

In both examples, the double structure seems to persist over most or
all of our $\sim$12-h observing sessions.  We can conclude that the
double structure is not an artifact of motion over short time scales.
Such rapid motion was in fact seen on 1997 January 16
\citep{Lebach+1999}.  Smaller motions on hour time scales occurred in
our 1998 March 1 and 1998 August 8 sessions, both of which show strong
flux-density variability \citep[see][]{Ransom-VLBA10th}.  These rapid
motions are much faster than the orbital motion, or that due to
rotation of the primary star, which are expected to be only
$\sim$0.03~mas over 3~hours, and thus not easily detectable in a VLBI
session.

\subsection{Polarization}
\label{spol}

For the 2004 March 6 epoch, for which we performed a full polarization
calibration, we measured a linear polarization fraction for \HR\ of
$<1$\%.  This is consistent with the expectation that Faraday rotation
within the corona of an RS~CVn system will substantially depolarize
any cm-wavelength radio emission \citep{Paredes2005}.  We do not
discuss linear polarization further.

\begin{figure}[ht]
\centering
\includegraphics[width=0.49\textwidth]{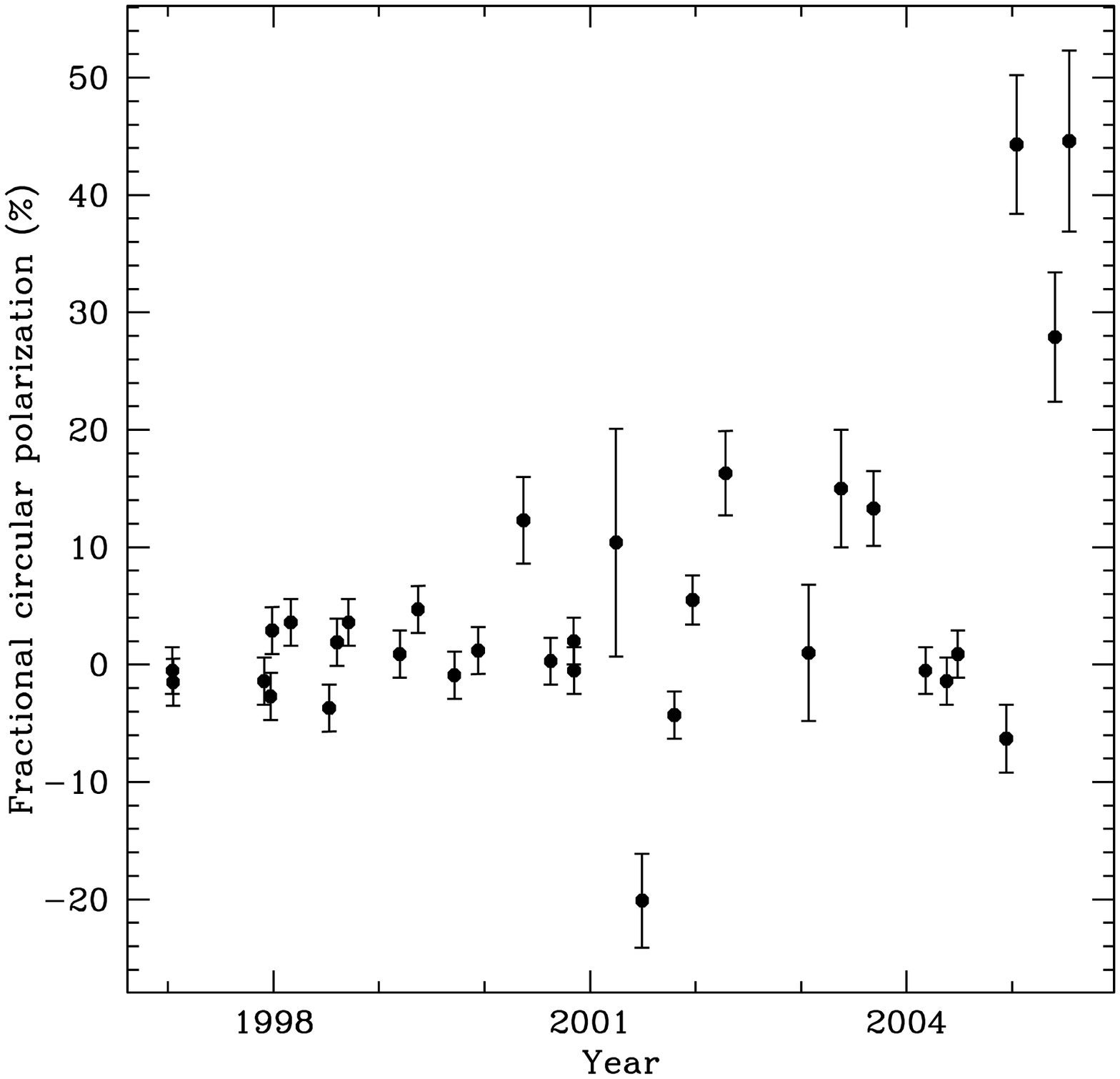}
\includegraphics[width=0.49\textwidth]{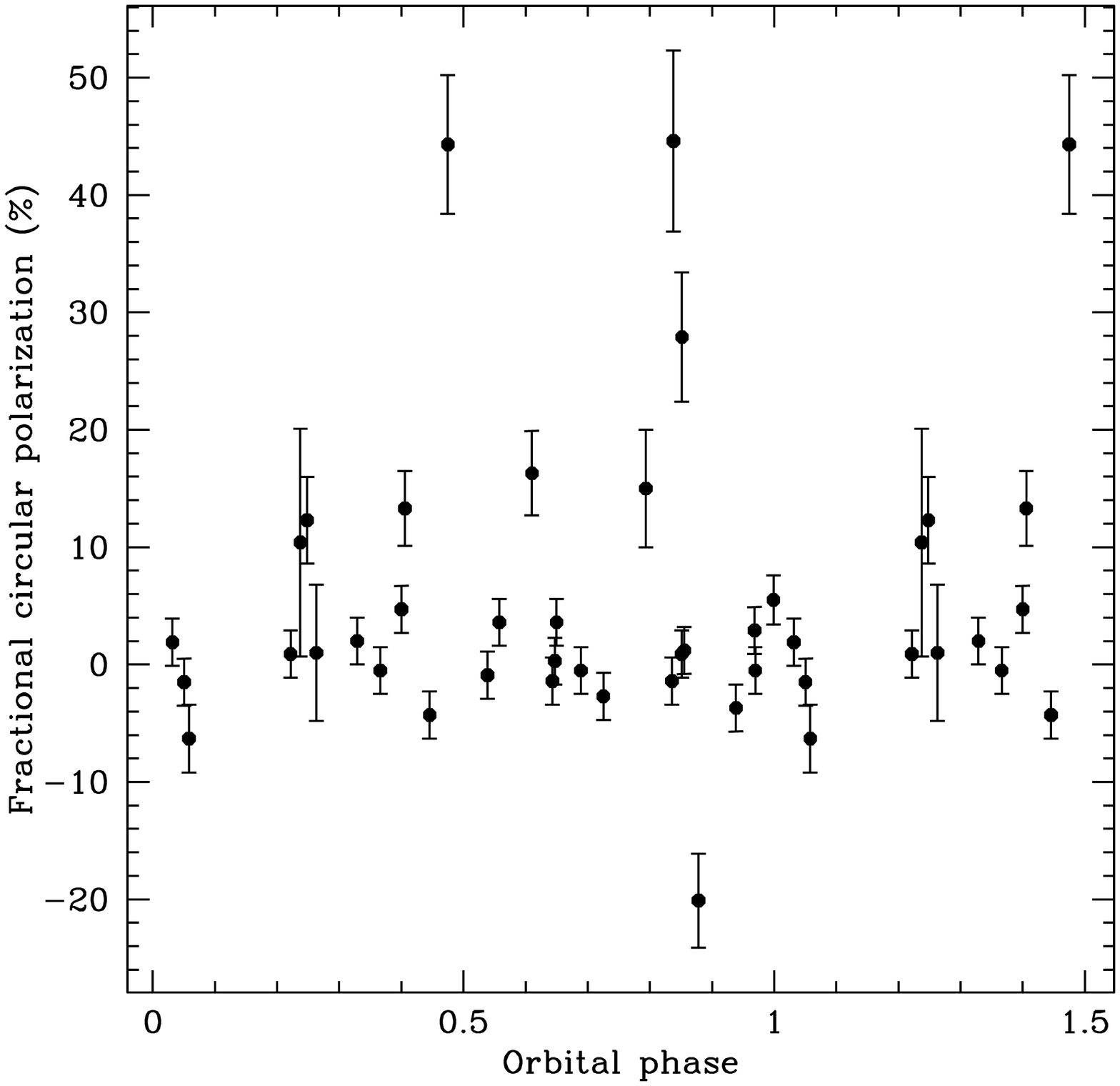}
\caption{Left panel: fractional circular polarization, $m_c$, (IEEE
convention), as determined from our VLA observations, plotted against
year.
Right panel: $m_c$ plotted against orbital phase \citep[using the
orbit of][]{Marsden+2005}.  Points with orbital phases between 0 and
0.5 are repeated on the right of the plot (with phases between 1.0 and
1.5) to make any possible cyclical variation more clearly visible.  }
\label{fvpolint}
\end{figure}

RS CVn systems are, on the other hand, expected to show substantial
circular polarization due to the gyro-synchrotron mechanism.  We did
therefore determine the average circular polarization for each
observing session from our VLA data (see Table~\ref{tepochs}).  We
plot the measured values of the fractional circular polarization,
$m_c$ in the left panel of Figure~\ref{fvpolint}. The maximum observed
value of $|m_c|$ was $\sim$46\% (on 2005 July 16).  Values of $|m_c| >
10$\% were only seen when the flux density was below 2~mJy.  Higher
values of $|m_c|$ are observed after 2004, when the total flux density
was generally lower, suggesting that $|m_c|$ is anti-correlated with
the total flux density.

Over the 32 observing sessions for which we had VLA data, the weighted
average of $m_c$ was $1.3 \pm 0.4$\%, with the standard deviation
being 13\%, and the cited uncertainty being statistical only.  This
result suggests a marginally significant ($3.1\sigma$) average
positive circular polarization.  However, an additional systematic
error, which we estimate at 1\% must be added because of the possible
deviations from the assumed value of zero for $m_c$ of our calibrator
sources as well as the uncorrected leakage terms.  The average value
of $m_c$ therefore cannot be regarded as significantly different from
0.

We also found no obvious correlation of $m_c$ with orbital phase: We
plot the fractional circular polarization against orbital phase in the
right panel of Figure~\ref{fvpolint}; no obvious correlation with
orbital phase is observed.  Note that here and throughout this paper,
we calculate the orbital phase at the midpoint of each observing
session, in fractions of an orbit period, as determined from the orbit
of \citet{Marsden+2005}. 

Although, as noted, the average value of $m_c$ was not significantly
different from 0, we did find evidence for a significantly positive
value of $m_c$ when the flux density was low.  We examine only those
epochs which had a flux density of $<2$~mJy, and found that the
weighted average $m_c$ was $4.3 \pm 0.9$\%.  Of these epochs, 10 out
of 12 have positive $m_c$.  The chance probability of observing this
many or more positive values of $m_c$ if the sign were random is
1.9\%.  We conclude that for low flux densities, there is a high
probability that \HR\ is right (IEEE convention) circularly polarized.

Significant circular polarization is commonly detected in RS~CVn
stars.  \HR\ seems to follow the trends observed for other systems, in
that $m_c$ has a somewhat consistent sign, but the magnitude decreases
with increasing flux density.  It has been proposed
\citep[e.g.,][]{Lestrade+1988, Mutel+1985} that this pattern is due to
flares radiating via the gyro-synchrotron mechanism.  Each flare
represents some release of energy.  Shortly after the initial release,
the emission is optically thick, hence of high brightness but of low
circular polarization.  As the flare decays, the brightness as well as
the optical thickness decrease, and the fractional circular
polarization increases.  The low-level ``quiescent'' emission might
then just be the superposition of the decays of many small flare
events.
Note that an alternative explanation for the quiescent emission which
has been proposed is that it represents gyro-synchrotron emission from
a thermal (Maxwellian) population of electrons
\citep[e.g.,][]{DrakeSL1989}.  Arguments against this hypothesis are
given in \citet{BeasleyG2000} and \citet{Paredes2005}.  Our
observation of high circular polarization when \HR's flux density was
low, as well as our observation of large source sizes, also when
\HR\ was quite weak (e.g., during the observing session of 2005 May
15) argues further against this hypothesis.  We therefore think that
the quiescent radio emission is due to a non-thermal, rather than a
thermal, population of electrons.

\begin{figure}[ht]
\centering
\includegraphics[width=\textwidth]{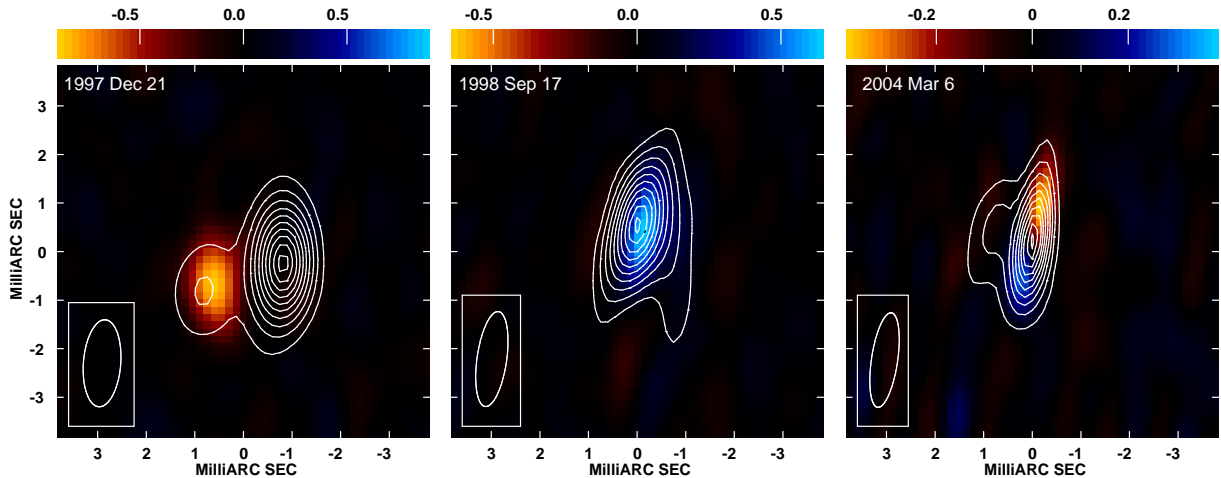}
\caption{Three example images showing the circularly polarized flux
density on the dates indicated.  The color scale shows the circularly
polarized flux density (Stokes $V$; IEEE convention) and is labeled
in m\Jb.  The contours show the total intensity (Stokes $I$), and are
drawn at $-10$, 10, 20, \dots\ 90, 98\% of the peak brightness, which
is 0.032, 0.013 and 0.010 m\Jb, respectively, for the three images.  
As in Figure~\ref{fimages}, the origin in each panel is the center of
the primary star as determined from the astrometric results in
\citepalias{GPB-V}. North is up and east is to the left. }
\label{fvpolimg}
\end{figure}

For our VLBI observing sessions, we recorded both right and left
circular polarization, and can therefore image the circularly
polarized flux density.  Figure~\ref{fvpolimg} shows three example
images.  The distribution of circular polarization clearly varies from
session to session.  Some sessions, such as that of 1998 September 17,
show a fractional circular polarization that is approximately constant
across the emission region; for other sessions, for example 2004 March
6, there are prominent gradients.

\subsection{Parametrization of the Emission Geometry}
\label{sparam}

\begin{figure}[ht]
\centering
\includegraphics[width=0.6\textwidth]{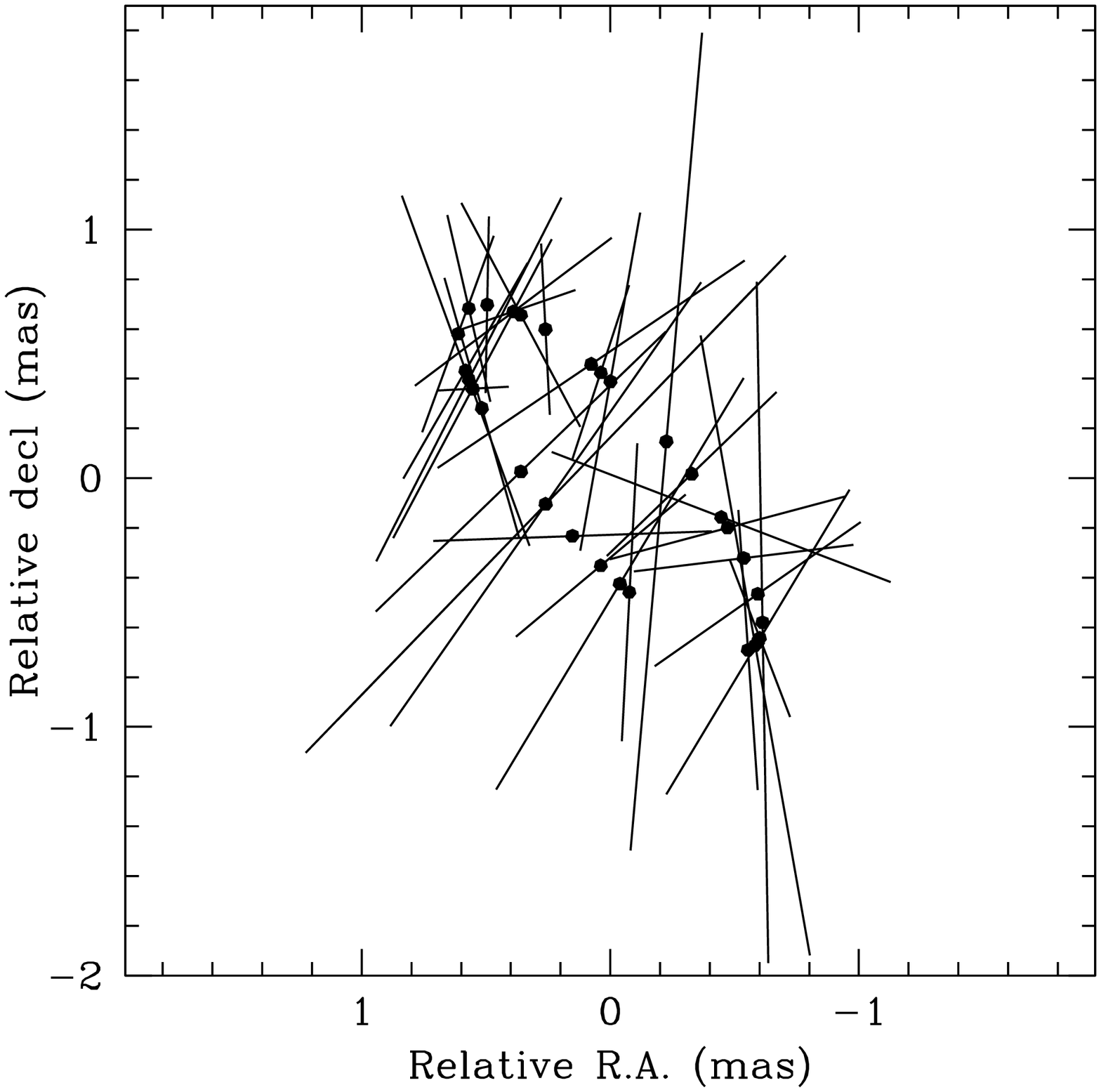}
\caption{The elongation of the radio emission region of \HR\ as a
function of position in the orbit.  The dots indicate the projected
position in the orbit of the primary star for each of our 35 observing
sessions, and the corresponding lines represent the FWHM major axis of
an elliptical Gaussian fit to the radio emission of that session.  The
origin of the coordinate system is the center of the orbit.}
\label{fpaorbit}
\end{figure}

To investigate the temporal evolution of the emission geometry, we
sought to parametrize it.  As a simple first approach
fitted an elliptical Gaussian to the image.  Although such a
fit does not adequately describe the complexity of the emission
region, especially when the emission region shows a double-structure,
the fitted major axis and its position angle nonetheless give an
indication of the overall extension and the direction of elongation of
the emission region.  We denote the vector major axis of the fitted
Gaussian by \psvb, and its
p.a.\ by \thelg.  Note that to avoid the bias introduced by
convolution with the CLEAN beam, we do not use directly the values
from the elliptical Gaussian fitted to the image, but rather those
from the ``deconvolved'' elliptical Gaussian.  The deconvolved
elliptical Gaussian is the one which, when convolved with the
elliptical Gaussian clean beam, results in the fitted elliptical
Gaussian.  The ``deconvolved'' fitted values of $|\psvb|$, \thelg, and
the axis ratio are given in Table~\ref{timages} for each session.  We
also plot \psvb\ on the projection of the binary orbit in
Figure~\ref{fpaorbit}.

Figure~\ref{fpaorbit} shows that the values of \thelg\ are not
randomly distributed, but rather show a preferred orientation.  The
average\footnote{Note that the quantities \psvb\ are axial, rather
than true, vectors in that $\thelg = 90$\arcdeg\ is equivalent to
$\thelg = -90$\arcdeg.  In order to meaningfully average \psvb, we follow
\citet{Batschelet1981} and first double the p.a.'s (\thelg), then
perform the usual vector average, and then halve the p.a.\ of the
result to obtain the final average value of \psvb.} 
of \psvb\ was 0.71~mas along \thelg = 157\arcdeg. A bootstrap
calculation gives a statistical uncertainty of 7\arcdeg\ for the
average direction of \psvb; however, systematic uncertainties, for
example stemming from the ``deconvolution'' of the clean beam, are
likely to be several times larger.  The probability\footnote{The
chance probability was estimated from a 5000-trial Monte-Carlo
simulation, with each trial using 35 vectors with the same lengths as
the measured values of \psvb, but with random orientations between
0\arcdeg\ and 180\arcdeg. }
of $n = 35$ random values being so well aligned is $\lesssim 1$\%.  We
note that the above calculation used a p.a.\ derived from a single
elliptical Gaussian fit to the entire emission region.

\section{Discussion}
\label{sdiscuss}

Using VLBI, we have produced a series of high-quality images of the
radio emission from the RS CVn star, \HR.  Our east-west resolution is
approximately equal to the stellar radius of the primary; our
north-south resolution is $\sim$3 times poorer.  The images show that
the radio emission exhibits a variety of morphologies, ranging from
being largely unresolved to having a clear double and on one occasion
(2003 Dec.\ 6) a possibly triple structure.  Although no other RS~CVn
system has been so extensively observed with VLBI, the images of other
systems are generally within the range of the morphologies we observed
for \HR.  The radio emission is highly variable and during most of our
VLBI observing sessions the total flux density varied significantly.
On occasion, time-resolved images show variations in morphology over
several hours. As we have mentioned, apparent motion of \HR's emission
region, correlated with evolution of its flux density on time scales of
one hour, is definitely seen in our 1997 Jan 16 session
\citep{Lebach+1999}, and may be present in at least two of our other
observing sessions.

What can be said about the nature of the variability of \HR's radio
emission?  Since \HR\ is a known binary, and its orbit is well known
from optical spectroscopy \citep{Marsden+2005, BerdyuginaIT1999a} and
from our VLBI astrometry (Papers IV and V), we next investigate
whether the radio emission varies significantly with the orbital phase
of the binary or in a secular fashion.

\subsection{Does the Radio Brightness Vary with Orbital Phase?}

\begin{figure}
\centering
\includegraphics[width=0.49\textwidth]{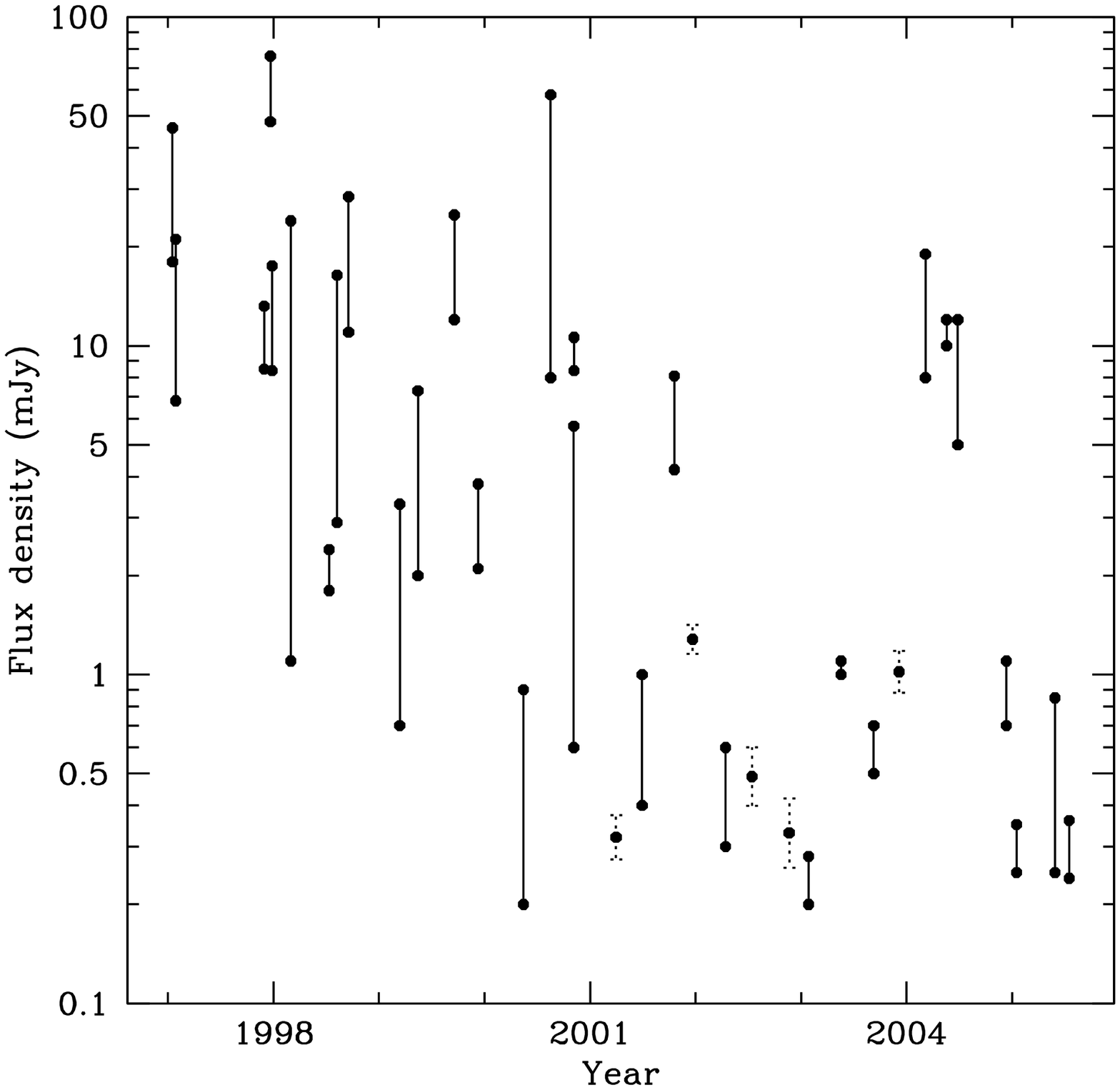}
\includegraphics[width=0.49\textwidth]{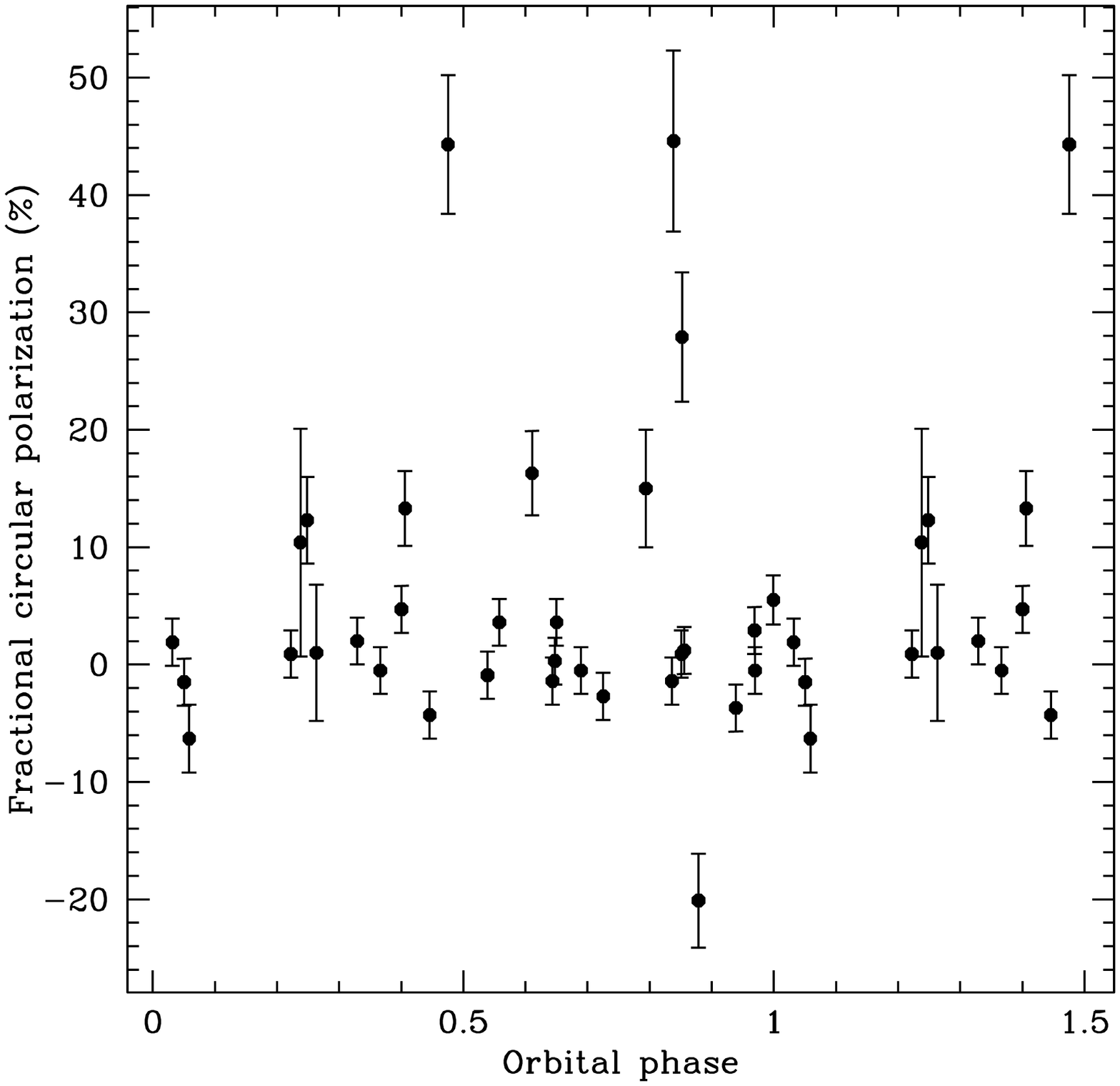}
\caption{Left panel: the logarithm of the 8.4-GHz total flux density
of \HR\ in milli-Janskies, plotted against time.  The total flux
density often varied within an observing session, and the range of
this variation is shown by the vertical lines that connect dots placed
at the extreme values of the flux densities during the observing
session.  In each case where the total flux density did not vary
discernibly within the observing interval or where the flux density
was determined from VLBI rather than VLA observations, the
observational uncertainty ($1\sigma$) is shown by a dotted error
bar centered on the dot; the fractional observational uncertainties for
the other sessions are similar.
Right panel: the same flux densities, but plotted against orbital
phase \citep[using the orbital parameters of][]{Marsden+2005}.  Points
with orbital phases between 0 and 0.5 are repeated on the right of the
plot (with phases between 1.0 ad 1.5) to make any possible cyclical
variation more clearly visible.  }
\label{ftotflx}
\end{figure}

In Fig.~\ref{ftotflx} we plot flux density against time and
also against orbital phase.  There is a secular decline in the flux
density, with lower flux densities being observed after late
1999.
No correlation with orbital phase is apparent.  In particular, we find
that the times when the flux density is high, presumably because of
outbursts, do not occur during any particular part of the orbit.  This
result is similar to that found for Algol \citep{Mutel+1998}, but
unlike that found for the RS~CVn binary HR~1099, where radio outbursts
seem to occur predominantly between the orbital phases of 0.50 to 0.67
\citep{SleeWR2008}.

\subsection{Orientation and Elongation of the Radio Emission Region}
\label{sorient}

As we showed in the previous section, no correlation between the total
flux density and the orbital phase is apparent.  Since the orbits are
nearly circular \citep{Marsden+2005, BerdyuginaIT1999a, Olah+1998},
one might not expect such a correlation if neither eclipsing nor
beaming effects are significant.  However, the emission region is
often extended by as much as the separation between the two stars
($\sim$2~mas), which leads to the question: is there any correlation
between the geometry of the emission region and the orbital phase?
For example, a correlation of the amount and/or angle of elongation of
the radio emission with orbital phase might be expected if the radio
emission originates predominantly along the line between the two stars
\citep[as is seen in the Algol system,][although we note that, unlike
\HR, Algol is an interacting system for which emission predominately
from within the region between the two stars is physically more
plausible]{Peterson+2010}.  To explore this question, we make use of
our Gaussian fits to the emission geometry from \S~\ref{sparam} and
Table~\ref{timages}, in particular, making use of the p.a., \thelg, of
the Gaussian fit to the images.

\begin{figure}
\centering
\includegraphics[width=0.49\textwidth]{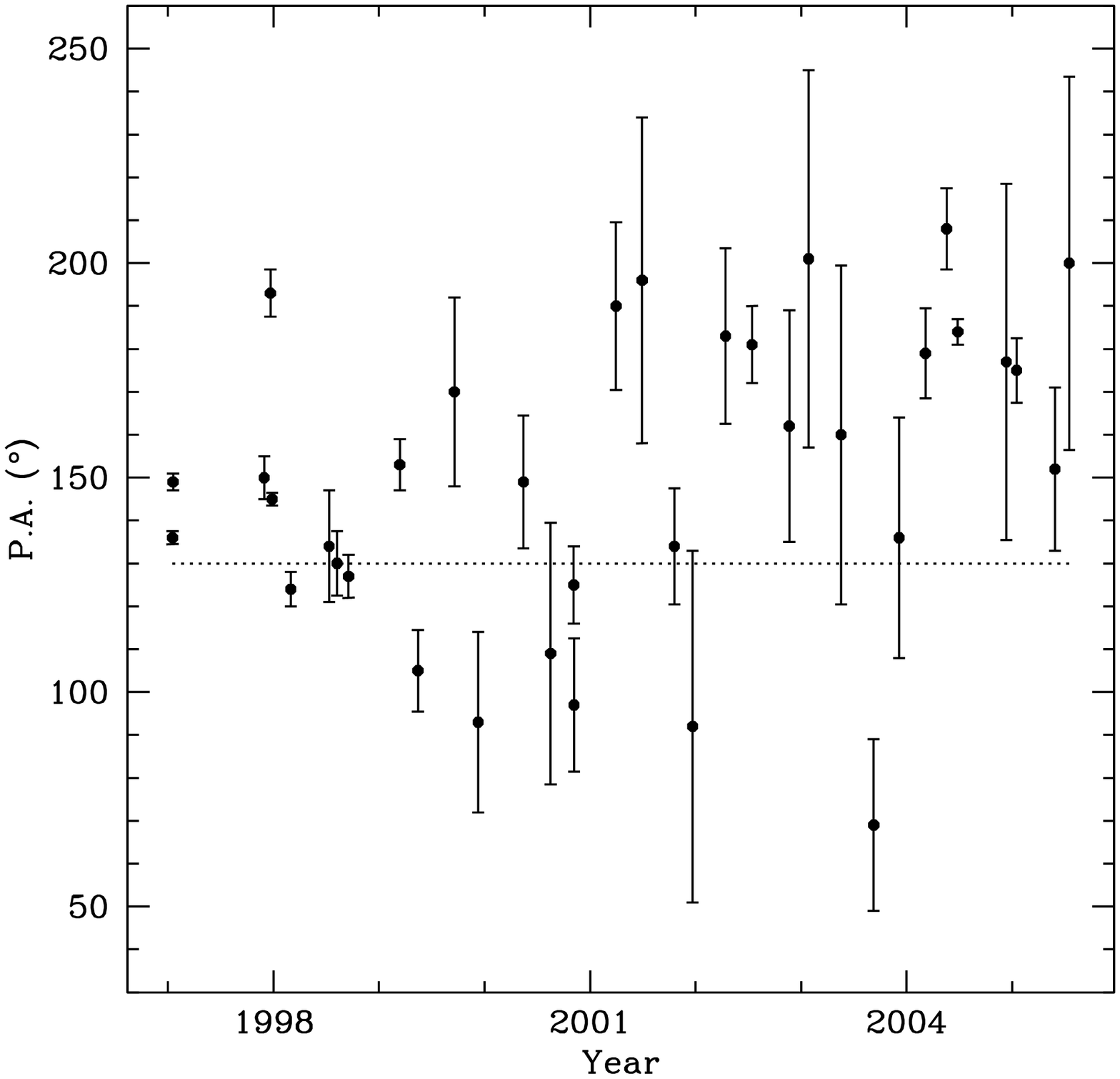}
\includegraphics[width=0.49\textwidth]{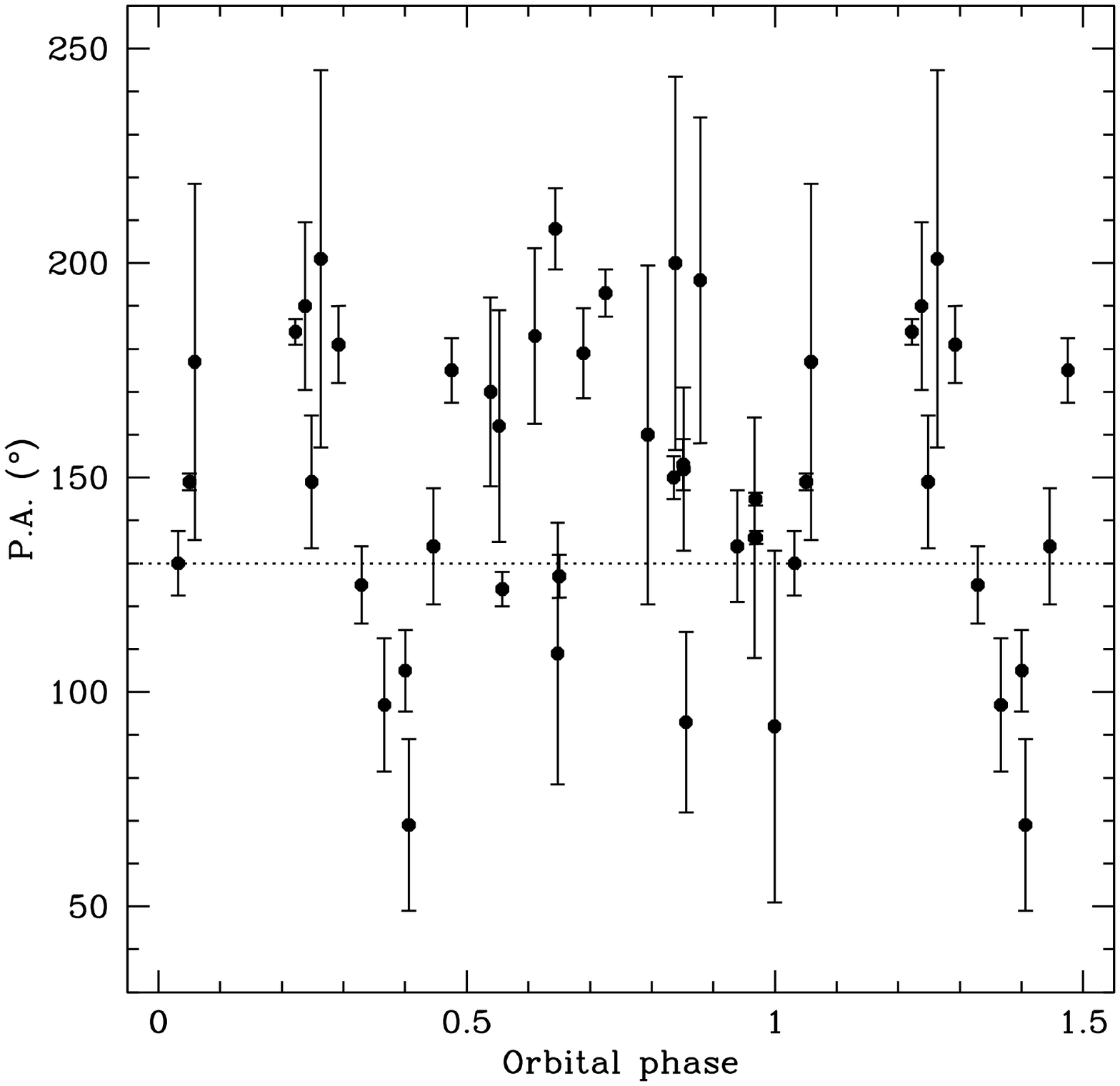}
\caption{Left panel: the p.a.\ of elongation of the radio emission
region, \thelg, determined from the VLBI images of \HR, plotted
against time.  We chose to plot a range of $\pm90$\arcdeg\ in p.a.\
around that of the orbital angular momentum as projected on the sky
\citepalias{GPB-VI}, which is indicated by a dotted
horizontal line (note that \thelg\ is axial: it can take on different possible
values only over a range of 180\arcdeg).  Right panel: the same
p.a.'s, but plotted against orbital phase \citep[using the orbit
of][]{Marsden+2005}.  Points with orbital phases between 0 and 0.5 are
repeated on the right of the plot (with phases between 1.0 and 1.5) to
make any possible cyclical variation more clearly visible.}
\label{ftheflg}
\end{figure}

Geometrical and astrophysical considerations suggest that the most
likely direction is either along the line from the primary to the
secondary or parallel to the orbit normal or perhaps the star's
rotation axis if it were different from the orbit normal.  In the
first case, one would expect \thelg\ would vary with the orbital
phase, while in the second, one would expect \thelg\ to remain
constant In Fig.~\ref{ftheflg} we plot \thelg\ against time and
against orbital phase.  Neither hypothesis is well supported.  Until
approximately 1999.5, \thelg\ is fairly well determined, with the data
possibly suggesting a secular variation with \thelg\ decreasing from
$\sim 140\arcdeg$ in 1995.0 to $\sim 100$\arcdeg\ by 1999.5.
Thereafter, \thelg\ is more poorly determined, largely because the
total flux densities and hence signal-to-noise ratios were lower due
to the aforementioned secular decrease in the average flux density.

In Figure \ref{fpaorbit} the direction of elongation of the radio
emission region does not show a correlation with the line joining the
primary to the secondary, implying that the radio emission region does
not stretch between the two stars. Since we think it unlikely that the
radio emission is associated with the secondary, we do not consider
that possibility further.  Since we find no evidence that the radio
emission is spatially associated with the line joining the two stars,
we therefore conclude that the radio emission is associated only with
the primary.  Indeed, this conclusion is consistent with our
astrometry \citepalias{GPB-V, GPB-VI}, which suggests that the radio
emission is associated with the primary.  We discuss the possibility
that the radio emission is associated with starspots on the primary
below in \S~\ref{slong}.

As noted in \S~\ref{sparam}, there is a significant tendency for the
emission region to be preferentially extended along p.a.\ =
157\arcdeg, which direction is reasonably close to that of the
sky-projected orbital angular momentum, which is at p.a.\ $130 \pm
13$\arcdeg\ \citepalias{GPB-V}\footnote{As we mention in
\S~\ref{sparam} above, the systematic uncertainty on the average angle
of elongation of the radio emission likely dominates the statistical
one of $\pm7$\arcdeg.  We can therefore not make a more definitive
statement as to whether the average direction of elongation of the
radio emission lies along the orbit normal within the uncertainties or
not.  We note that any bias in our estimate of the elongation of the
emission region towards the average elongation of the restoring beam
at p.a.\ = 174\arcdeg\ would bring it closer to the angle of the orbit
normal of $130 \pm 13$\arcdeg.}
The radio emission, therefore, seems to extend preferentially in the
projected direction close to that of the orbital angular momentum.
This alignment suggests that the emission tends to be associated with
the magnetic poles of the star, provided the magnetic poles are near
the rotational ones.  This location is consistent with the polar-cap
model proposed for the Algol system by \citet{Mutel+1998}.

The average extent of the emission region (as given by the fitted
Gaussian FWHM major-axis length, $b$) was $1.4 \pm 0.4$~mas,
corresponding at 96~pc to $(1.9 \pm 0.6) \times10^{12}$~cm, or $1.1
\pm 0.3$ times the star's diameter
\citep[for which we again use the value of
$\sim$27~\Rsol\ from][]{BerdyuginaIT1999a}.  This result suggests
that the bulk of the radio emission generally occurs within an area
not much larger in size than the stellar disk.

\subsection{Rapid Evolution of the Radio Structure}

By dividing an observing session temporally, we can assess any
possible rapid evolution of the image geometry.  A rapid change in the
apparent position accompanied by a large change in flux density is
seen in several sessions.  The most dramatic example was the already
mentioned change in apparent position of $\sim$0.9~mas over a period
of 1.4~hr on 1997 January 16 \citep[see][]{Lebach+1999}.
Such a change in apparent position could be due either to (1) fast
motion at $\sim 1000$~\kms\ of flare-energized electrons in a single
magnetic-loop structure, or (2) two spatially distinct components
separated by $\sim$1~mas, with a rapid change in their relative
brightness.  Given the elongation of the emission region for the 1997
Jan.\ observing session (see Fig.~\ref{fimages}), the latter is more
plausible.  However, for the 1998 March and 1998 August sessions, in
which the emission region is quite compact, motion of the energized
electrons is a likely possibility.

We see double (and in one case, apparent triple) structure in
$\sim$25\% of our images.  Images made from temporal subsets of some
of our observing sessions, for example those shown in
Figure~\ref{f1997dec}, show that the double structure usually persists
over periods $>6$~hr.  We found in \S~\ref{sorient} above that, on
average, the extent of the radio emission on the sky was comparable to
the size of the stellar disk.  On occasion though, the extent of the
radio emission (as given by the FWHM of an elliptical Gaussian fitted
to the image) is up to about twice the stellar diameter\footnote{We
note that the largest value for a FWHM major axis of the radio
emission was observed on 2005 Jan 15 and was 3.3~mas, corresponding
to 2.5 stellar diameters.  That particular value, however, is rather
uncertain as \HR\ was very weak for that session.  Of the
better-determined major-axis values, the largest one corresponds to
$\sim2.2\times$ the diameter of the stellar disk.}.
It seems plausible therefore, that most of the radio emission occurs
not far from the stellar surface, but on occasion, relatively bright
radio emission is generated at distances of up a stellar diameter
above the surface.

It is likely that the radio emission originates in active regions,
e.g., magnetic-loop structures, approximately a stellar radius in size
\citep[see, e.g.,][]{Peterson+2010, Mullan+2006, Franciosini+1999,
Lestrade+1988}, which are rooted on the surface of the star.  Indeed,
\HR\ is of the spectral type K2, and \citet{Mullan+2006} show that the
largest coronal loops, with sizes up to two stellar radii, occur in
stars of type K2 or later.
Although the radio brightness of each emission region is often
variable on time scales of $\sim$1~hr or less, the regions themselves
appear longer lived, since the radio emission does not generally seem
to move on scales of the stellar radius over periods of a few hours.
The rotation and orbital motion of the primary will produce proper
motions of $\sim$10~{\mbox{$\mu$as\,hr$^{-1}$}}, which is below what
can be reliably determined from our images.  The lack of motions on
the short time scales of the flux-density variations suggest that more
rapid bulk or pattern motion, for example due to plasma moving along
the magnetic-loop structures, are not prevalent or at least generally
occur on scales smaller than the stellar radius.

\subsection{Fourier Analysis of Quantities Derived from Images}

The near-equality between the photometric period of \HR\
\citep[variable between 23.8 and 25.2 days; see][]{Strassmeier+1997}
with the spectroscopic one of 24.64877~d suggests that the rotation of
\HR\ primary is tidally locked to its orbital
motion (note that, given the variable nature of the dark spots on the
surface of the star, an exact equivalence between the photometric and
spectroscopic periods is not expected).  Both the lack of any clear
correlations between either the location of the emission relative to
the center of the star, or the direction of elongation of the emission
region with the orbital phase, and the observations of apparent motion
on hour time scales, therefore suggest that the brightest emission
does not always emanate from a region centered on a single spot on the
rotating surface of the star.

Optical Doppler imaging of \HR's surface has shown that there are dark
spots, which cover $\gtrsim$15\% of the star's surface
\citep{BerdyuginaM2006, Berdyugina+2000}. These dark spots are
presumably the cause of the photometric variability.  The photometric
period, however, does not correspond precisely to the orbital one:
The photometric period is variable, ranging between $\sim$24 and
$\sim$25~d between 1993 and 1999
\citep{Strassmeier+1997, Berdyugina+2000}.  In fact, multi-band
photometry can be used to constrain the spot population independent of
Doppler imaging, and gives generally consistent results
\citep{Zellem+2010}.
Over the period 1996 to 1999, the dark spots responsible for the
photometric variation occur preferentially at two active
longitudes\footnote{The stellar longitude is defined by assuming
tidally-locked rotation, with longitude zero being for the 
meridian facing the secondary.}
on opposite hemispheres (in longitude), but seem to drift in longitude
by about 0.05\arcdeg~d$^{-1}$, resulting in a photometric period
which can vary slightly from the spectroscopic one.

\begin{figure}
\centering
\includegraphics[width=0.99\textwidth]{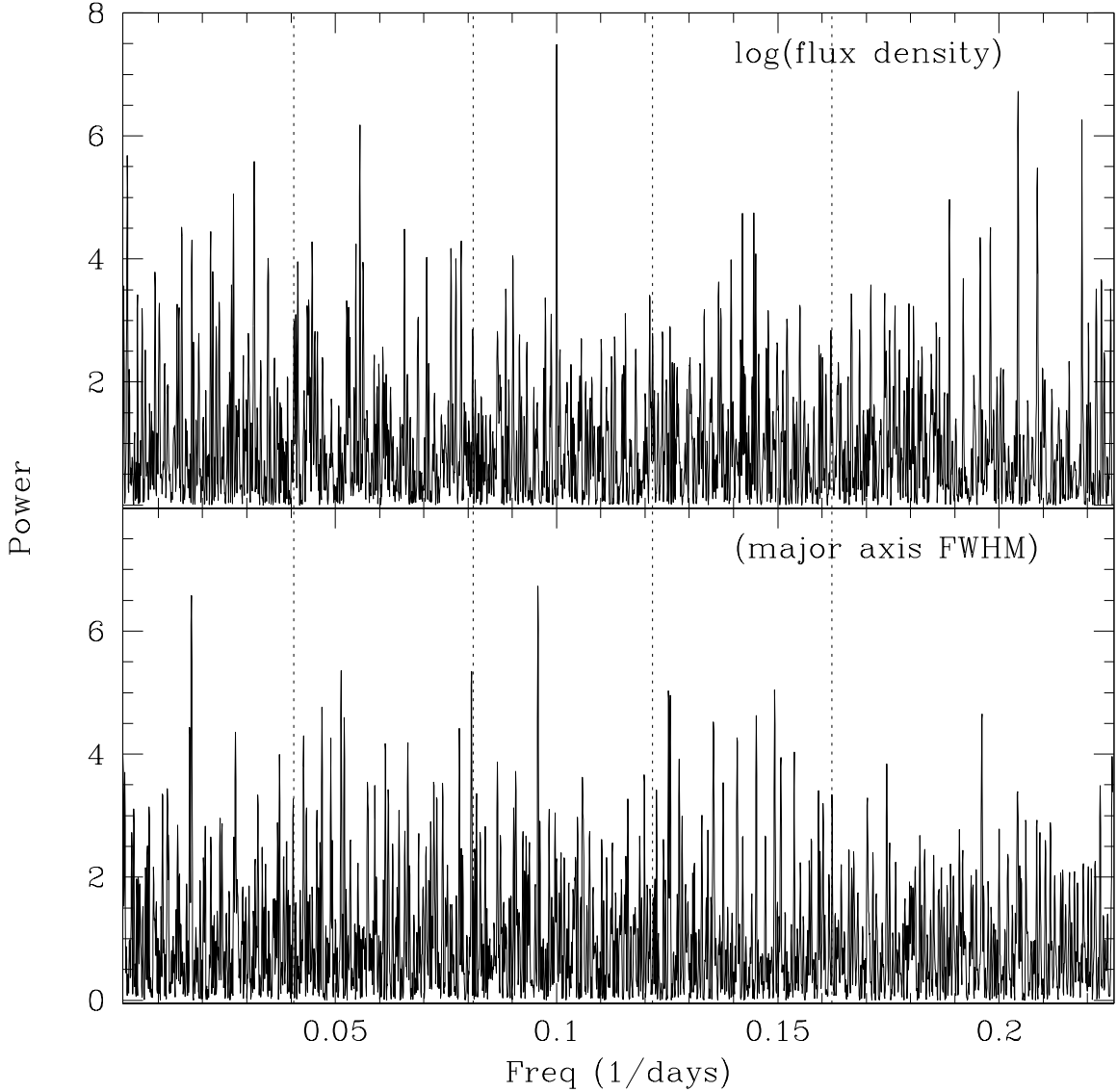}
\caption{Upper panel: normalized Lomb-Scargle periodogram of the log
of the total flux density of \HR\ derived from our 35 observing
sessions of observations.  The dotted lines indicate the spectroscopic
binary frequency of (1/24.64877)~d$^{-1}$ and its first three
harmonics.  Lower panel: a normalized Lomb-Scargle periodogram of the
maximum angular extent of the radio emission, as parametrized by the
FWHM major axis of an elliptical Gaussian fitted to the radio images.}
\label{fspect}
\end{figure}

Do our measurements show evidence of periodicities other than the
orbital one?  To identify any possible periodicities in the radio
data, we produced normalized Lomb-Scargle periodograms
\citep{Scargle1982} of several quantities derived from the radio
observations, namely the total flux density, the fractional variation
of the flux density within the observing session, the cosine and sine
of the p.a.\ of elongation, the major axis length of an elliptical
Gaussian fit to the radio images and the average (over an observing
session) degree of circular polarization.  In none of these
periodograms did we find any significant periodicity. All the
peaks seen have chance probabilities of 30\% or higher
\citep[probabilities are calculated according to][]{Scargle1982}.
As an example, we show in Fig.~\ref{fspect} the periodograms for the
total flux density and the maximum angular extent of the radio
emission as parametrized by the FWHM major axis of an elliptical
Gaussian fit to each radio image.  No peaks higher than would be
expected by chance are observed, and the highest peaks observed in
either periodogram are not near the orbital frequency or its first
three harmonics.  No significant peaks at longer periods, such as
the 294~d (frequency = $\sim$0.0035 d$^{-1}$) Rieger-like flaring
periodicity found in the RS~CVn system UX~ARI \citep{Massi2007,
MassiNC2005}, are observed.  We note, however, that there are local
maxima near some of the harmonics of the orbital frequency, indicating
that there may well be some periodic behavior, but not at a level
reliably distinguishable with our data from mere random variation.
In particular, there is moderately high peak near the second harmonic
in the periodogram for the major axis length, at a frequency of
0.0808~\invd, or $1.99 \times$ the orbital one of 0.040570~\invd.
This peak is the fourth-highest in the periodogram.  If the location
of the periodogram peaks were random, the chance that one of the four
highest peaks would fall within 1\% of the first four harmonics of the
orbital frequency would be $\sim$3\%.  If the periodicity is
associated with the projection onto the sky of some dimension which
rotates with the stars in their orbit, then geometrical considerations
in fact suggest a frequency {\em double} the orbital one, as is
observed.  We also note that the photometric period seems to be
slightly different than the orbital one.  \citet{Berdyugina+2000}
found a spot rotation frequency of 0.997 times the orbital one during
the period 1996 to 1999; its second harmonic would be very close to
the periodogram peak in question.

In conclusion, no statistically significant periodicities are visible
in the characteristics of the radio emission: its total flux density,
its short time scale variability, or its spatial extent.  No
significant periodicities are present in the astrometric residuals
either \citepalias[see][]{GPB-V}.  We note however, that there is
evidence, albeit inconclusive, for some geometrical effect on the
extent of the radio emission region associated with either the
rotation of the binary, or more likely the rotation of the starspots
on the primary star.

\subsection{Comparison of Radio Images with Optical Doppler Imaging}
\label{slong}

Optical spectra from short exposures of $\lesssim20$~min allow the imaging
of portions of
the star's photosphere using Doppler imaging
\citep[e.g.,][]{BerdyuginaM2006, Berdyugina+2000, BerdyuginaIT1999a}.
As mentioned earlier, such imaging shows that the dark spots occur
primarily at one or more long-lived active longitudes on the surface
of the star.  \citet{RibarikOS2003} came to a similar conclusion by
analyzing the photometric data.  The optical observations showed that
the most intense spots usually occur on the side towards the
secondary, i.e., at a stellar longitude near 0, although, as noted
above, the active longitudes were found to change slowly with time
\citep{Marsden+2007, BerdyuginaM2006, RibarikOS2003, Berdyugina+2000},
likely as a result of differential rotation of the stellar surface,
as we see in the Sun, and as has been observed in a different RS~CVn
star, II Pegasi \citep{Roettenbacher+2010}.
The dark spots also seem more likely to occur at stellar latitudes
nearer the pole than the equator \citep{BerdyuginaM2006,
Berdyugina+2000}.

As our astrometric fit gives a good estimate of the expected position
of the center of the star for each observing session, can we calculate
the location of the brightness peak of the radio emission as projected
onto the star's surface for comparison with the location of the
optical dark spots?  Given that the height of the radio-emission peak
above the stellar surface is unknown, the stellar longitude and
latitude cannot be directly obtained.  However, as we showed in
\citetalias{GPB-VI}, the scatter in the positions of the peak
brightness points relative to the estimated center position of the
star was consistent with a distribution having a scale height of only
$\sim$0.2 stellar radii above the surface of the star.  We also
concluded there that ``spillover'' emission, that is where the
emission originates on the side of the star away from us, but enough
of it spreads over the limb to be detected, was responsible for only a
relatively small fraction of the observed radio brightness peaks.

\begin{figure}
\centering
\includegraphics[width=0.49\textwidth]{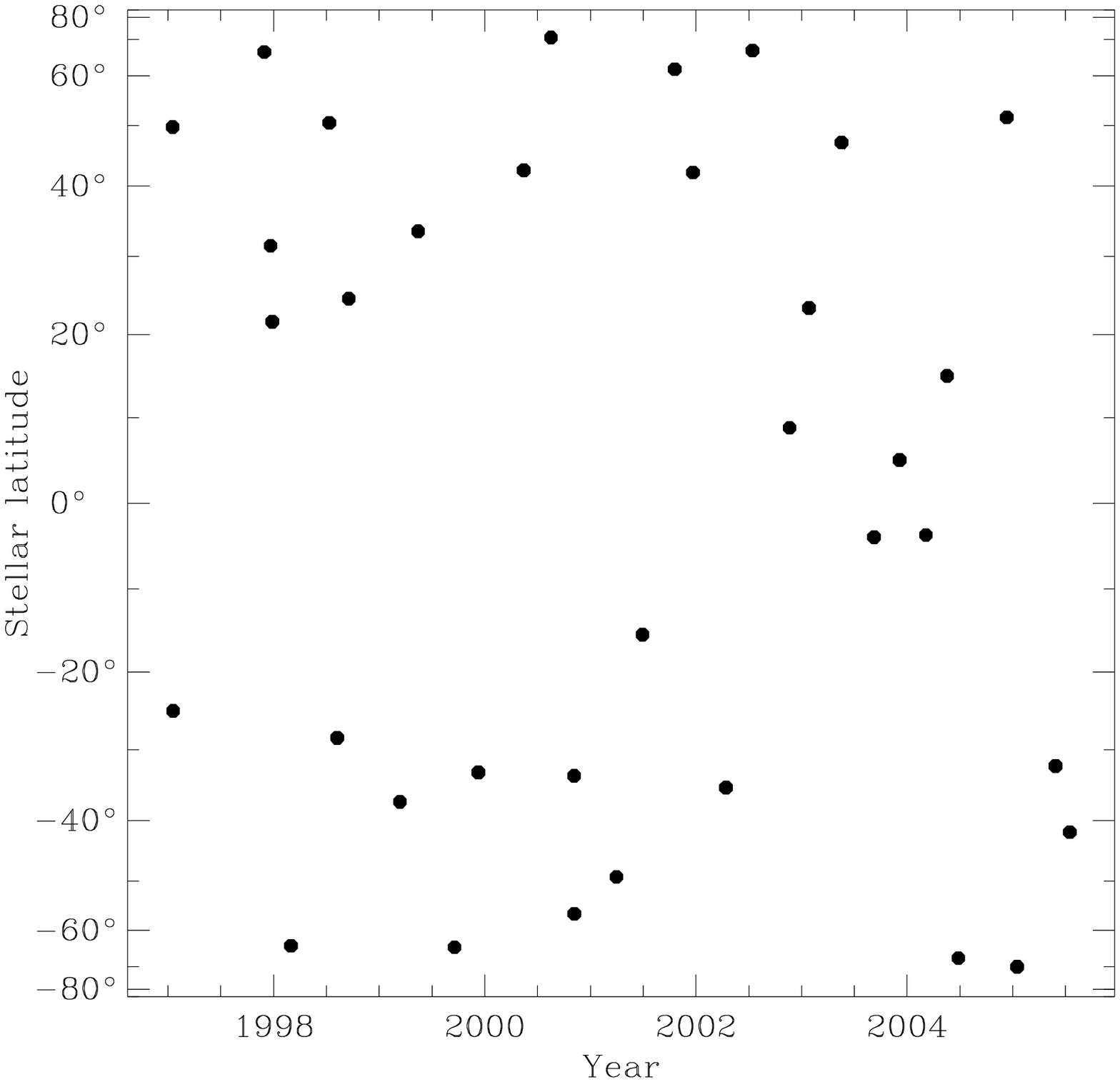}
\includegraphics[width=0.49\textwidth]{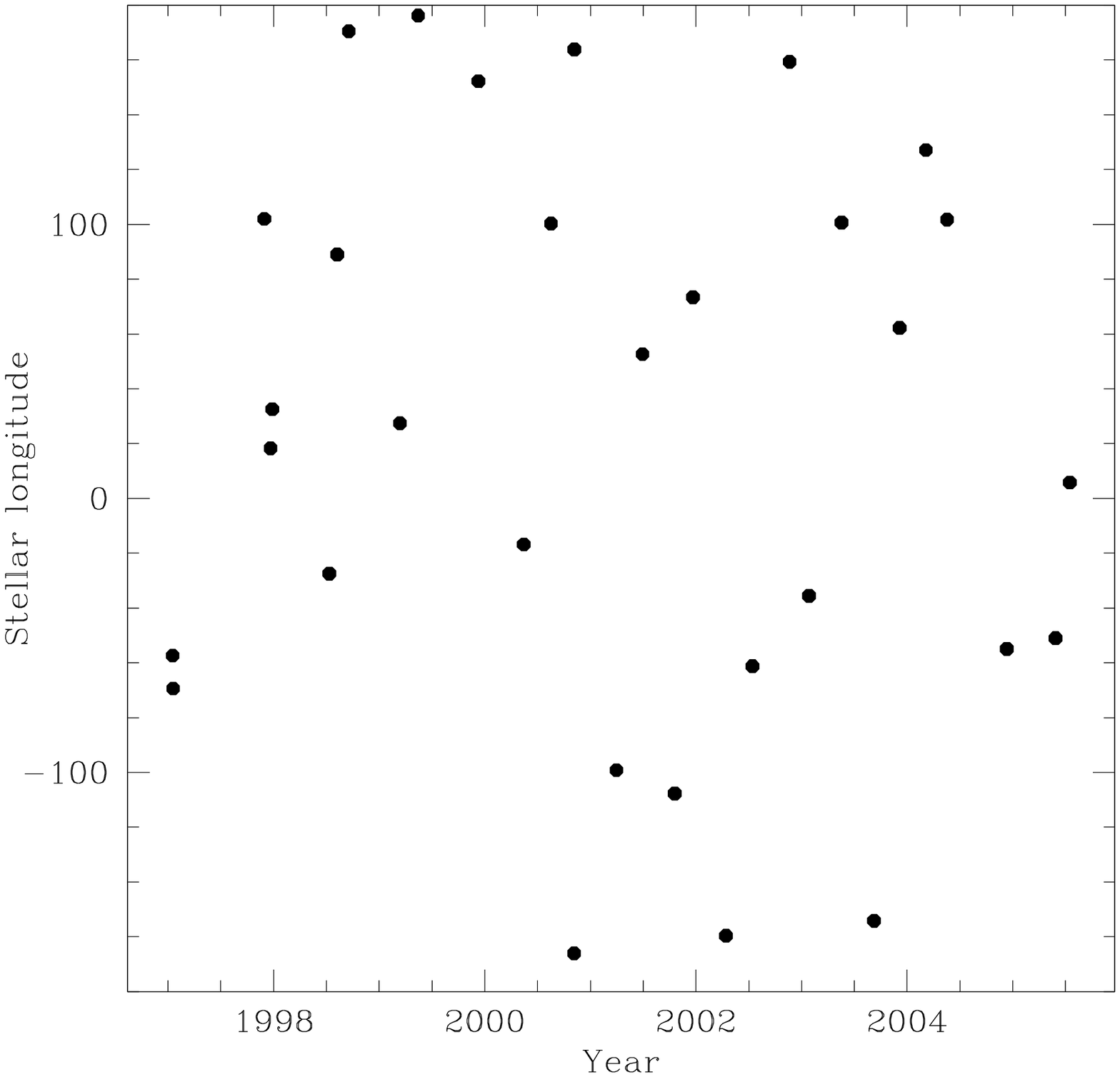}
\caption{The latitude and longitude on the surface of the star
corresponding to the peak brightness point of the radio emission. We
calculate the latitude and longitude by assuming the star's center to
be at the position given by our astrometric fit, and by taking the
star's angular radius to be 0.64~mas (see \S~\ref{slong} for the
assumptions used in calculating the stellar latitude and longitude).
Since the longitude becomes degenerate near the pole of the star, we
omit in the right panel the 4 points which were $<0.1$~mas from
the pole on the sky.}
\label{flatlong}
\end{figure}

We therefore make the simplifying approximations that the emission
peak is located exactly on the stellar surface and is confined to the
half of the surface facing us.  With these approximations, we can then
calculate stellar latitude and longitude of the emission peak for each
observing session except for those where the brightness peak falls
outside the stellar disk.  For those sessions, we calculate the
latitude and longitude for the nearest point on the limb of the star.
We again use the value of 13.3~\Rsol\ \citep{BerdyuginaIT1999a}, which
corresponds to 0.64~mas, for the radius of the star.  Although the
calculation of stellar latitude and longitude of the brightness peak
is approximate and could be seriously in error for a few sessions
because of the above simplifying assumptions, in general our estimates
should be adequate for our purpose of determining whether the emission
regions are preferentially associated with some parts of the star's
surface. We plot our values of the stellar latitude and longitude of
the brightness peak against time in Figure~\ref{flatlong}.

The radio emission shows no strong preference for any particular
stellar longitudes.  In particular, we find no significant
preponderance of radio emission near zero longitudes (see
Figs.~\ref{flatlong}).  Eliminating indeterminate longitudes
associated with high latitudes, we are left with 31 values for the
longitude of the brightness peak.  Of these, 17, or $55\pm12$\%, have
longitudes in the range $-90$\arcdeg\ to 90\arcdeg (where longitude
zero faces the secondary), which is consistent with the spot longitude
being random.

As we already found in \citetalias{GPB-VI}, the radio brightness peak
may show a preference for stellar latitudes nearer the poles.
For example, $8\pm 3$ points have a latitude either $\ge \pm
60$\arcdeg or $\le \pm -60$\arcdeg, whereas only 5 would be expected
by chance.

The optical dark spots also show a preference for latitudes near the
poles \citep{Marsden+2007, BerdyuginaM2006}, suggesting that there may
be a connection between the regions of bright radio emission and the
optical dark spots.  Although the localization on the star's surface
of the radio bright regions and the optical dark spots is not of
sufficient accuracy to warrant definitive conclusions, the evidence
suggests that the radio-bright regions do not directly correspond to
the optical dark spots, since the radio bright regions do not seem to
show the same preferred longitudes as do the dark spots.  Nonetheless,
the dark spots may represent the locations from which at least some
short-lived radio-bright flares emanate, as on the sun.

\subsection{The Nature of the Radio Emission}

Our radio data suggests that the radio emission is highly variable
temporally and spatially, and is at least on occasion, localized in
regions covering only a fraction of the stellar surface.  This picture
is supported by the observation of time-variability on scales much
shorter than the orbital period, and of rapid ($\lesssim 1$~hr)
positional variability accompanied by rapid changes in flux density.
In other words the radio emission seems likely related to some rapid
energy release localized on the stellar surface.

Indeed, the total radio emission is likely due to several distinct but
temporally overlapping energy releases, or flares, consistent with the
model proposed for RS~CVn radio emission by \citet{FranciosiniC1995,
Mutel+1985}.  The flares have time scales of only a few hours, with
individual flares originating at different locations on, or above, the
surface of the star. On occasion, more than one such spot is present
and we observe double, or possibly triple, structure.  The radio
emission is extended preferentially in a polar direction, perhaps
because of magnetic field lines extending also in that direction.  On
average the angular extent of the emission region is similar to that
of the stellar disk, although it can be larger.  This observation
suggests that the radio emission mostly occurs near the stellar
surface, but on occasion also occurs significantly above it.

Although the short time scales and the prevalence of multiple structure
suggest that the radio emission seems concentrated in small regions of
the star's surface, we have no evidence that the radio flare-spots are
associated with the optical dark spots on the stellar surface.  It
may, however, arise from extended magnetic-loop structures whose feet
are fixed to these active surface regions.  The occasional release of
magnetic energy stored at or below the surface in the long-lived spot
regions could energize electrons trapped in the loops, which then
radiate via the gyro-synchrotron mechanism.

The average flux density of \HR\ decreased during our 8.5-year observing
period.  During the first few years of our observing program, the
above-mentioned flares were more frequent, occurring sometimes within
just a few hours of one another (see Fig.~\ref{flightcurves}).  Toward
the later part of our program, flares were much less frequent,
separated perhaps by several days.  If there is a steady-state
population of radio-emitting electrons, our observations set a limit
on this quiescent emission at a level $\lesssim$0.3~mJy.

\section{Summary}

We discuss the unprecedented series of 35 VLBI radio images of the
RS~CVn star \HR\ at 8.4~GHz, taken over a period of 8.5~years. These
astrometric and imaging VLBI observations were undertaken because \HR\
was chosen as the guide star for \GPB\@. As well as the imaging VLBI
observations, we obtained total flux-density measurements using the
VLA\@.  Our astrometric solution for the star's proper motion, orbit
and parallax allows us to align the images in a star-centered frame.
We find that:

\begin{trivlist}

\item{1.} The radio emission from the star is rapidly variable.
  During most of our observing sessions, the total flux density varied
  by a factor of at least two over the course of several hours.

\item{2.} Significant circular polarization was observed for about
half of our observing sessions.  The fractional circular polarization
was also rapidly variable, and had a tendency to be higher when the
total flux density was lower.   For epochs with high circular
polarization and low flux density, the circular polarization
was predominately positive (IEEE convention).

\item{3.} The morphology of the radio emission is also variable.
Mostly the emission region has a single-peaked structure, but about
25\% of the time, we observed two (and on one occasion three) peaks.
On average, the emission region is elongated by $\sim 1.4 \pm 0.4$~mas
(FWHM).

\item{4.}  On average, the emission region is comparable in size to
the disk of the primary star.  It is also, on average, somewhat more
extended in the direction of the sky-projected angular momentum of the
binary orbit.  The radio emission is likely associated with one or
more regions on or above the surface of the star.

\item{5.} We searched for, but did not find, any sign of periodic
behavior, either at the known orbital period (24.64877~d), or at any
other periods. In particular, neither flux density, degree of
circular polarization, degree or angle of elongation of the emission
region showed any significant periodicities.

\item{6.} The regions of brightest radio emission show no preference
for any particular stellar longitudes, unlike the dark spots seen in
the optical.  Nonetheless, like the optical dark spots, the regions of
bright radio emission regions show a preference for stellar latitudes
near the poles.  Despite a likely relationship between the mechanisms,
the regions of brightest radio emission are short-lived, and do not
seem to be directly associated with the long-lived optical dark spots.

\item{7.} The radio emission may arise due to short-lived flares, with
each flare occupying only a small part of the stellar surface.  These
flares may be associated with a magnetic field which has its strongest
component along the direction of the rotational axis.

\end{trivlist}

\acknowledgements

ACKNOWLEDGMENTS.  This research was primarily supported by NASA,
through a contract with Stanford University to SAO, and a subcontract
from SAO to York University.  The National Radio Astronomy Observatory
(NRAO) is a facility of the National Science Foundation operated under
cooperative agreement by Associated Universities, Inc. The DSN is
operated by JPL/Caltech, under contract with NASA\@.  This research
has made use of the United States Naval Observatory (USNO) Radio
Reference Frame Image Database (RRFID)\@.  We have also made use of
NASA's Astrophysics Data System Abstract Service, developed and
maintained at SAO\@. Jeff Cadieux and Julie Tome helped with the data
reduction during their tenure as students at York University.
Finally, we thank the anonymous referee for useful comments.

\bibliography{gpb,gpb-temp,gpb-ours}

\begin{thebibliography}{51}
\expandafter\ifx\csname natexlab\endcsname\relax\def\natexlab#1{#1}\fi

\bibitem[{{Baars} {et~al.}(1977){Baars}, {Genzel}, {Pauliny-Toth}, \&
  {Witzel}}]{Baars+1977}
{Baars}, J.~W.~M., {Genzel}, R., {Pauliny-Toth}, I.~I.~K., \& {Witzel}, A.
  1977, \aap, 61, 99

\bibitem[{{Bartel} {et~al.}(2011){Bartel}, {Bietenholz}, {Lebach}, {Lederman},
  {Petrov}, {Ransom}, {Ratner}, \& {Shapiro}}]{GPB-III}
{Bartel}, N., {Bietenholz}, M.~F., {Lebach}, D.~E., {Lederman}, J.~I.,
  {Petrov}, L., {Ransom}, R.~R., {Ratner}, M.~I., \& {Shapiro}, I.~I. 2012,
  this issue (Paper III)

\bibitem[{{Bartel} {et~al.}(2008){Bartel}, {Ransom}, {Bietenholz}, {Lebach},
  {Ratner}, {Shapiro}, \& {Lestrade}}]{NB-GPB2008}
{Bartel}, N., {Ransom}, R.~R., {Bietenholz}, M.~F., {Lebach}, D.~E., {Ratner},
  M.~I., {Shapiro}, I.~I., \& {Lestrade}, J.-F. 2008, in IAU Symposium, Vol.
  248, IAU Symposium, ed. W.~J. {Jin}, I.~{Platais}, \& M.~A.~C. {Perryman},
  190--191

\bibitem[{{Batschelet}(1981)}]{Batschelet1981}
{Batschelet}, E. 1981, {Circular Statistics in Biology} (London: Academic
  Press)

\bibitem[{{Beasley} \& {G{\"u}del}(2000)}]{BeasleyG2000}
{Beasley}, A.~J., \& {G{\"u}del}, M. 2000, \apj, 529, 961

\bibitem[{{Berdyugina} {et~al.}(2000){Berdyugina}, {Berdyugin}, {Ilyin}, \&
  {Tuominen}}]{Berdyugina+2000}
{Berdyugina}, S.~V., {Berdyugin}, A.~V., {Ilyin}, I., \& {Tuominen}, I. 2000,
  \aap, 360, 272

\bibitem[{{Berdyugina} {et~al.}(1999{\natexlab{a}}){Berdyugina}, {Ilyin}, \&
  {Tuominen}}]{BerdyuginaIT1999b}
{Berdyugina}, S.~V., {Ilyin}, I., \& {Tuominen}, I. 1999{\natexlab{a}}, \aap,
  349, 863

\bibitem[{{Berdyugina} {et~al.}(1999{\natexlab{b}}){Berdyugina}, {Ilyin}, \&
  {Tuominen}}]{BerdyuginaIT1999a}
------. 1999{\natexlab{b}}, \aap, 347, 932

\bibitem[{{Berdyugina} \& {Marsden}(2006)}]{BerdyuginaM2006}
{Berdyugina}, S.~V., \& {Marsden}, S.~C. 2006, in Astronomical Society of the
  Pacific Conference Series, Vol. 358, Astronomical Society of the Pacific
  Conference Series, ed. R.~{Casini} \& B.~W. {Lites}, 385

\bibitem[{{Bietenholz} {et~al.}(2003){Bietenholz}, {Bartel}, \&
  {Rupen}}]{SN93J-3}
{Bietenholz}, M.~F., {Bartel}, N., \& {Rupen}, M.~P. 2003, \apj, 597, 374,
  arXiv:astro-ph/0307382

\bibitem[{{Briggs}(1995)}]{Briggs1995}
{Briggs}, D.~S. 1995, in Bulletin of the American Astronomical Society,
  Vol.~27, Bulletin of the American Astronomical Society, 1444

\bibitem[{{Briggs} {et~al.}(1999){Briggs}, {Schwab}, \&
  {Sramek}}]{BriggsSS1999}
{Briggs}, D.~S., {Schwab}, F.~R., \& {Sramek}, R.~A. 1999, in Astronomical
  Society of the Pacific Conference Series, Vol. 180, Synthesis Imaging in
  Radio Astronomy II, ed. G.~B. {Taylor}, C.~L. {Carilli}, \& R.~A. {Perley}
  (San Francisco, CA: ASP), 127

\bibitem[{{Drake} {et~al.}(1989){Drake}, {Simon}, \& {Linsky}}]{DrakeSL1989}
{Drake}, S.~A., {Simon}, T., \& {Linsky}, J.~L. 1989, \apjs, 71, 905

\bibitem[{{Dulk}(1985)}]{Dulk1985}
{Dulk}, G.~A. 1985, \araa, 23, 169

\bibitem[{{ESA}(1997)}]{PerrymanE1997}
{ESA}. 1997, {The HIPPARCOS and TYCHO catalogues,
  SP-1200} (Noordwijk, Netherlands: ESA)

\bibitem[{{Franciosini} \& {Chiuderi-Drago}(1994)}]{FranciosiniC1995}
{Franciosini}, E., \& {Chiuderi-Drago}, F. 1994, Radiophysics and Quantum
  Electronics, 37, 403

\bibitem[{{Franciosini} {et~al.}(1999){Franciosini}, {Massi}, {Paredes}, \&
  {Estalella}}]{Franciosini+1999}
{Franciosini}, E., {Massi}, M., {Paredes}, J.~M., \& {Estalella}, R. 1999,
  \aap, 341, 595

\bibitem[{{Hall}(1976)}]{Hall1976}
{Hall}, D.~S. 1976, in ASSL Vol. 60: IAU Colloq. 29: Multiple Periodic Variable
  Stars, ed. W.~S. {Fitch} (Dordrecht: Reidel), 287

\bibitem[{{Hjellming}(1988)}]{Hjellming1988}
{Hjellming}, R.~M. 1988, in Galactic and Extragalactic Radio Astronomy, ed.
  G.~L. Kellermann, K.~I.~\&~Verschuur (Berlin and New York: Springer-Verlag),
  381--438

\bibitem[{{Jones} {et~al.}(1996){Jones}, {Brown}, {Stewart}, \&
  {Slee}}]{Jones+1996}
{Jones}, K.~L., {Brown}, A., {Stewart}, R.~T., \& {Slee}, O.~B. 1996, \mnras,
  283, 1331

\bibitem[{{Lebach} {et~al.}(2011){Lebach}, {Bartel}, {Bietenholz}, {Campbell},
  {Gordon}, {Lederman}, {Lestrade}, {Ransom}, {Ratner}, \& {Shapiro}}]{GPB-IV}
{Lebach}, D.~E. {et~al.} 2012, this issue (Paper IV)

\bibitem[{{Lebach} {et~al.}(1999){Lebach}, {Ratner}, {Shapiro}, {Ransom},
  {Bietenholz}, {Bartel}, \& {Lestrade}}]{Lebach+1999}
{Lebach}, D.~E., {Ratner}, M.~I., {Shapiro}, I.~I., {Ransom}, R.~R.,
  {Bietenholz}, M.~F., {Bartel}, N., \& {Lestrade}, J.-F. 1999, \apjl, 517, L43

\bibitem[{{Lestrade} {et~al.}(1995){Lestrade}, {Jones}, {Preston}, {Phillips},
  {Titus}, {Kovalevsky}, {Lindegren}, {Hering}, {Froeschle}, {Falin},
  {Mignard}, {Jacobs}, {Sovers}, {Eubanks}, \& {Gabuzda}}]{Lestrade+1995}
{Lestrade}, J.-F. {et~al.} 1995, \aap, 304, 182

\bibitem[{{Lestrade} {et~al.}(1988){Lestrade}, {Mutel}, {Preston}, \&
  {Phillips}}]{Lestrade+1988}
{Lestrade}, J.-F., {Mutel}, R.~L., {Preston}, R.~A., \& {Phillips}, R.~B. 1988,
  \apj, 328, 232

\bibitem[{{Marsden} {et~al.}(2007){Marsden}, {Berdyugina}, {Donati}, {Eaton},
  \& {Williamson}}]{Marsden+2007}
{Marsden}, S.~C., {Berdyugina}, S.~V., {Donati}, J.-F., {Eaton}, J.~A., \&
  {Williamson}, M.~H. 2007, Astronomische Nachrichten, 328, 1047

\bibitem[{{Marsden} {et~al.}(2005){Marsden}, {Berdyugina}, {Donati}, {Eaton},
  {Williamson}, {Ilyin}, {Fischer}, {Mu{\~n}oz}, {Isaacson}, {Ratner}, {Semel},
  {Petit}, \& {Carter}}]{Marsden+2005}
{Marsden}, S.~C. {et~al.} 2005, \apjl, 634, L173

\bibitem[{{Massi}(2007)}]{Massi2007}
{Massi}, M. 2007, Memorie della Societa Astronomica Italiana, 78, 247

\bibitem[{{Massi} \& {Aaron}(1999)}]{MassiA1999}
{Massi}, M., \& {Aaron}, S. 1999, \aaps, 136, 211

\bibitem[{{Massi} {et~al.}(1988){Massi}, {Felli}, {Pallavicini}, {Tofani},
  {Palagi}, \& {Catarzi}}]{Massi+1988}
{Massi}, M., {Felli}, M., {Pallavicini}, R., {Tofani}, G., {Palagi}, F., \&
  {Catarzi}, M. 1988, \aap, 197, 200

\bibitem[{{Massi} {et~al.}(2005){Massi}, {Neidh{\"o}fer}, {Carpentier}, \&
  {Ros}}]{MassiNC2005}
{Massi}, M., {Neidh{\"o}fer}, J., {Carpentier}, Y., \& {Ros}, E. 2005, \aap,
  435, L1, arXiv:astro-ph/0503257

\bibitem[{{Mullan} {et~al.}(2006){Mullan}, {Mathioudakis}, {Bloomfield}, \&
  {Christian}}]{Mullan+2006}
{Mullan}, D.~J., {Mathioudakis}, M., {Bloomfield}, D.~S., \& {Christian}, D.~J.
  2006, \apjs, 164, 173

\bibitem[{{Mutel} {et~al.}(1985){Mutel}, {Lestrade}, {Preston}, \&
  {Phillips}}]{Mutel+1985}
{Mutel}, R.~L., {Lestrade}, J.~F., {Preston}, R.~A., \& {Phillips}, R.~B. 1985,
  \apj, 289, 262

\bibitem[{{Mutel} {et~al.}(1998){Mutel}, {Molnar}, {Waltman}, \&
  {Ghigo}}]{Mutel+1998}
{Mutel}, R.~L., {Molnar}, L.~A., {Waltman}, E.~B., \& {Ghigo}, F.~D. 1998,
  \apj, 507, 371

\bibitem[{{Olah} {et~al.}(1998){Olah}, {Marik}, {Houdebine}, {Dempsey}, \&
  {Budding}}]{Olah+1998}
{Olah}, K., {Marik}, D., {Houdebine}, E.~R., {Dempsey}, R.~C., \& {Budding}, E.
  1998, \aap, 330, 559

\bibitem[{{Paredes}(2005)}]{Paredes2005}
{Paredes}, J.~M. 2005, in EAS Publications Series, ed. L.~I. {Gurvits},
  S.~{Frey}, \& S.~{Rawlings}, 187--206

\bibitem[{{Peterson} {et~al.}(2010){Peterson}, {Mutel}, {G{\"u}del}, \&
  {Goss}}]{Peterson+2010}
{Peterson}, W.~M., {Mutel}, R.~L., {G{\"u}del}, M., \& {Goss}, W.~M. 2010,
  \nat, 463, 207

\bibitem[{{Peterson} {et~al.}(2011){Peterson}, {Mutel}, {Lestrade},
  {G{\"u}del}, \& {Miller Goss}}]{Peterson+2011}
{Peterson}, W.~M., {Mutel}, R.~L., {Lestrade}, J.-F., {G{\"u}del}, M., \&
  {Miller Goss}, W. 2011, ArXiv e-prints, 1104.5005

\bibitem[{{Ransom} {et~al.}(2011{\natexlab{a}}){Ransom}, {Bartel},
  {Bietenholz}, {Lebach}, {Lederman}, {Luca}, {Ratner}, \& {Shapiro}}]{GPB-II}
{Ransom}, R.~R., {Bartel}, N., {Bietenholz}, M.~F., {Lebach}, D.~E.,
  {Lederman}, J.~I., {Luca}, P., {Ratner}, M.~I., \& {Shapiro}, I.~I.
  2012{\natexlab{a}}, this issue (Paper II)

\bibitem[{{Ransom} {et~al.}(2011{\natexlab{b}}){Ransom}, {Bartel},
  {Bietenholz}, {Lebach}, {Lestrade}, {Ratner}, \& {Shapiro}}]{GPB-VI}
{Ransom}, R.~R., {Bartel}, N., {Bietenholz}, M.~F., {Lebach}, D.~E.,
  {Lestrade}, J.-F., {Ratner}, M.~I., \& {Shapiro}, I.~I. 2012{\natexlab{b}},
  this issue (Paper VI)

\bibitem[{{Ransom} {et~al.}(2002){Ransom}, {Bartel}, {Bietenholz}, {Lebach},
  {Ratner}, {Shapiro}, \& {Lestrade}}]{Ransom+2002}
{Ransom}, R.~R., {Bartel}, N., {Bietenholz}, M.~F., {Lebach}, D.~E., {Ratner},
  M.~I., {Shapiro}, I.~I., \& {Lestrade}, J.-F. 2002, \apj, 572, 487

\bibitem[{{Ransom} {et~al.}(2003){Ransom}, {Bartel}, {Bietenholz}, {Ratner},
  {Lebach}, {Shapiro}, \& {Lestrade}}]{Ransom+2003}
{Ransom}, R.~R., {Bartel}, N., {Bietenholz}, M.~F., {Ratner}, M.~I., {Lebach},
  D.~E., {Shapiro}, I.~I., \& {Lestrade}, J.-F. 2003, \apj, 587, 390,
  astro-ph/0301413

\bibitem[{{Ransom} {et~al.}(2005){Ransom}, {Bartel}, {Bietenholz}, {Ratner},
  {Lebach}, {Shapiro}, \& {Lestrade}}]{Ransom-VLBA10th}
{Ransom}, R.~R., {Bartel}, N., {Bietenholz}, M.~F., {Ratner}, M.~I., {Lebach},
  D.~I., {Shapiro}, I.~I., \& {Lestrade}, J.-F. 2005, in Astronomical Society
  of the Pacific Conference Series, Vol. 340, Future Directions in High
  Resolution Astronomy, ed. J.~{Romney} \& M.~{Reid}, 506

\bibitem[{{Ratner} {et~al.}(2011){Ratner}, {Bartel}, {Bietenholz}, {Lebach},
  {Lestrade}, {Ransom}, \& {Shapiro}}]{GPB-V}
{Ratner}, M.~I., {Bartel}, N., {Bietenholz}, M.~F., {Lebach}, D.~E.,
  {Lestrade}, J.-F., {Ransom}, R.~R., \& {Shapiro}, I.~I. 2012, this issue
  (Paper V)

\bibitem[{{Rayner} {et~al.}(2000){Rayner}, {Norris}, \& {Sault}}]{RaynerNS2000}
{Rayner}, D.~P., {Norris}, R.~P., \& {Sault}, R.~J. 2000, \mnras, 319, 484

\bibitem[{{Rib{\'a}rik} {et~al.}(2003){Rib{\'a}rik}, {Ol{\'a}h}, \&
  {Strassmeier}}]{RibarikOS2003}
{Rib{\'a}rik}, G., {Ol{\'a}h}, K., \& {Strassmeier}, K.~G. 2003, Astronomische
  Nachrichten, 324, 202

\bibitem[{{Roettenbacher} {et~al.}(2010){Roettenbacher}, {Harmon},
  {Vutisalchavakul}, \& {Henry}}]{Roettenbacher+2010}
{Roettenbacher}, R.~M., {Harmon}, R.~O., {Vutisalchavakul}, N., \& {Henry},
  G.~W. 2010, ArXiv e-prints, 1009.2308

\bibitem[{{Scargle}(1982)}]{Scargle1982}
{Scargle}, J.~D. 1982, \apj, 263, 835

\bibitem[{{Shapiro} {et~al.}(2011){Shapiro}, {Bartel}, {Bietenholz}, {Lebach},
  {Lestrade}, {Ransom}, \& {Ratner}}]{GPB-I}
{Shapiro}, I.~I., {Bartel}, N., {Bietenholz}, M.~F., {Lebach}, D.~E.,
  {Lestrade}, J.-F., {Ransom}, R.~R., \& {Ratner}, M.~I. 2012, this issue,
  (Paper I)

\bibitem[{{Slee} {et~al.}(2008){Slee}, {Wilson}, \& {Ramsay}}]{SleeWR2008}
{Slee}, O.~B., {Wilson}, W., \& {Ramsay}, G. 2008, \pasa, 25, 94, 0802.0819

\bibitem[{{Strassmeier} {et~al.}(1997){Strassmeier}, {Bartus}, {Cutispoto}, \&
  {Rodono}}]{Strassmeier+1997}
{Strassmeier}, K.~G., {Bartus}, J., {Cutispoto}, G., \& {Rodono}, M. 1997,
  \aaps, 125, 11

\bibitem[{{Zellem} {et~al.}(2010){Zellem}, {Guinan}, {Messina}, {Lanza},
  {Wasatonic}, \& {McCook}}]{Zellem+2010}
{Zellem}, R., {Guinan}, E.~F., {Messina}, S., {Lanza}, A.~F., {Wasatonic}, R.,
  \& {McCook}, G.~P. 2010, \pasp, 122, 670

\end{thebibliography}

\end{document}